\documentclass{aa}
\usepackage{amsmath}
\usepackage{graphicx}
\usepackage{breqn} 
\usepackage[mathscr]{euscript} 

\usepackage[utf8]{inputenc}

\usepackage{booktabs}
\usepackage{adjustbox}

\usepackage{tablefootnote}
\usepackage{caption}
\usepackage{subcaption}
\usepackage{ulem}

\usepackage{multicol}

\usepackage{multirow}

\usepackage{hyperref}

\usepackage{natbib} 
\bibpunct{(}{)}{;}{a}{}{,} 

\usepackage{txfonts}
\graphicspath{{Graphics/}}

\begin{document}

   \title{Supercritical colliding wind binaries}
   \author{Leandro Abaroa\inst{1,2}\thanks{leandroabaroa@gmail.com}, 
          Gustavo E. Romero\inst{1,2}, 
           \and Pablo Sotomayor\inst{1,2}
           }

   \offprints{Leandro Abaroa}
  \institute{Instituto Argentino de Radioastronomía, CICPBA-CONICET-UNLP \\                                                                                       
   Villa Elisa, La Plata, Argentina                                                             
   \and Facultad de Cs. Astron\'omicas y Geof\'{\i}sicas, Universidad Nacional de La Plata \\                                                                           
   Paseo del Bosque S/N (1900), La Plata, Argentina 
                }

   \date{Received / Accepted}


\abstract
{Particle-accelerating colliding-wind binaries (PACWBs) are systems that are formed by two massive and hot stars and produce nonthermal radiation. The key elements of these systems are fast winds and the shocks that they create when they collide. Binaries with nonaccreting young pulsars have also been detected as nonthermal emitters, again as a consequence of the wind--wind interaction. Black holes might produce nonthermal radiation by this mechanism if they accrete at super-Eddington rates. In such cases, the disk is expected to launch a radiation-driven wind, and if this wind has an equatorial component, it can collide with the companion star yielding a PACWB. These systems are supercritical colliding wind binaries.}
{We aim to characterize the particle acceleration and nonthermal radiation produced by the collision of winds in binary systems composed of a superaccreting black hole and an early-type star.}
{We estimated the terminal velocity of the disk-driven wind by calculating the spatial distribution of the radiation fields and their effect on disk particles. We then found the location of the wind collision region and calculated the timescales of energy gain and losses of relativistic particles undergoing diffusive particle acceleration. With this information, we were able to compute the associated spectral energy distribution of the radiation. We calculated a number of specific models with different parameters to explore this scenario.}
{We find that the interaction of winds can produce nonthermal emission from radio up to tens of GeV, with luminosities in the range of $\sim 10^{33}$--$10^{35} \, {\rm erg \, s^{-1}}$, which for the most part are contributed  by electron synchrotron and inverse Compton radiation.}
{We conclude that supercritical colliding wind binaries, such as some ultraluminous X-ray sources and some Galactic X-ray binaries, are capable of accelerating cosmic rays and producing nonthermal electromagnetic emission from radio to $\gamma$-rays, in addition to the thermal components.}

\authorrunning{L. Abaroa et al.}
\titlerunning{Super critical colliding wind binaries}

\maketitle
\section{Introduction}

Early-type stars are very hot and their radiation fields can launch powerful particle winds \citep{1999isw..book.....L}. Such winds quickly reach supersonic velocities and accelerate to terminal velocities in the range $(2-4)\times10^{3}$ km s$^{-1}$ \citep{1978ApJ...225..893A, 2012A&A...537A..37M}. When two massive stars with powerful winds form a binary system, the winds collide producing shocks separated by a contact discontinuity from where matter is evacuated \citep[e.g.,][]{1992ApJ...386..265S}. A reverse shock moves in the wind of each star. When such shocks are adiabatic, they can accelerate suprathermal particles up to relativistic energies \citep{1993ApJ...402..271E, 2020MNRAS.495.2205P}. These particles, in turn, cool mainly by synchrotron radiation and inverse Compton upscattering of stellar photons, emitting nonthermal radiation \citep{1993ApJ...402..271E, 2003A&A...399.1121B, 2006ApJ...644.1118R, 2007A&ARv..14..171D,2014ApJ...789...87R,2016A&A...591A.139D,2021MNRAS.504.4204P}. Proton acceleration can also lead to gamma-ray emission through $pp$ collisions and the subsequent $\pi^0$ decays \citep[e.g.,][]{2017A&A...603A.111B, 2019ApJ...871...55G}.  

The actual fraction of particle-accelerating colliding-wind binaries (PACWBs) among massive colliding wind binaries (CWBs) is not well known. \cite{2013A&A...558A..28D}  list 43 confirmed cases, mostly detected at radio wavelengths. These authors mention several other candidates, 
and new sources have been found since the publication of this latter work \citep[e.g.,][]{2015A&A...579A..99B,2016A&A...591A.139D}. The total kinetic power of these systems ranges from $\sim 10^{34}$ to more than $10^{37} \, \rm{erg \, s^{-1}}$. The most extreme cases are WR89, WR98, and WR140, with powers of between 6 and 8 times $10^{37} \, \rm{erg \, s^{-1}}$. Less than $10^{-7} $ of this power is finally radiated through  synchrotron radio emission. The most luminous nonthermal radio-emitting CWB is WR140, with a total radio luminosity of $\sim 2.6\times 10^{30} \, \rm{erg \, s^{-1}}$. 

Contrary to the radio emission, high-energy radiation has been more difficult to detect in CWBs. At X-rays, the thermal component usually dominates and hinders the detection of nonthermal components. In the gamma-ray domain, only two systems have been detected so far: $\eta$ Carinae and WR11. The latter is the nearest known CWB. At $d \sim 340$ pc, it shows a gamma-ray luminosity in the \textit{Fermi}-LAT energy range of $L_{\gamma} = (3.7 \pm 0.7) \times 10^{31} \, \rm{erg \, s^{-1}}$. This luminosity amounts to $\sim 6 \times 10^{-6}$ of the total wind kinetic power \citep{2016MNRAS.457L..99P}. Similar fractions for other, more distant PACWBs yield fluxes that are  undetectable with the currently available instrumentation. The notable exception is the mentioned  $\eta$ Carinae. 

$\eta$ Carinae is a heavily obscured and peculiar object. The system includes a luminous blue variable (LBV) star of about 90 solar masses and a secondary Wolf-Rayet (WR) star of $\sim 30$ solar masses. $\eta$ Carinae is the most luminous binary in the Galaxy, with a bolometric luminosity of about $5 \times 10^6$ $L_{\odot}$. The mass-loss rate of the primary is extremely high, reaching up to $10^{-3}$ $M_{\odot} \, \rm{yr}^{-1}$. The binary was detected in hard X-rays by {\it INTEGRAL} \citep{2008A&A...477L..29L} and {\it Suzaku} \citep{2008MNRAS.388L..39O}, suggesting the presence of relativistic electrons in the system. {\it AGILE} detected gamma rays from $\eta$ Carinae for the first time \citep{2009ApJ...698L.142T}. The system was subsequently detected by {\it Fermi} \citep{2010ApJ...723..649A} with a luminosity of $\sim 10^{34} \, \rm{erg \, s^{-1}}$. The observations reveal the presence of a hard component in the spectrum around periastron, which disappears near apastron. Such a component has been explained through the decay of $\pi^0$ produced by relativistic protons   interacting with the dense stellar wind \citep{2011A&A...526A..57F}. There is a clear variability with the orbital phase. Different behaviors are observed at low ($0.3 - 10$ GeV) and high ($> 10$ GeV) gamma-ray energies. The low-energy component is likely produced by inverse Compton scattering of stellar photons \citep{2017A&A...603A.111B}.

The case of $\eta$ Carinae suggests that super-Eddington systems might be particularly powerful PACWBs. When a compact object such as a black hole accretes with rates that exceed the Eddington rate, the radiation pressure on the surface of the disk will overcome the gravitational attraction and matter will be expelled from the surface of the disk in the form of a strong wind. Such winds can rival and even surpass those of the most luminous CWBs in terms of kinetic power. When the donor star is a hot early-type star also endowed with a wind, a supercritical colliding wind binary (SCWB) can be formed. Such systems should have strong shocks and are potential particle accelerators and nonthermal emitters. 

In our Galaxy, there are some examples of black hole X-ray binaries with disks that launch strong outflows. Two examples are GRS 1915+105 \citep{1994Natur.371...46M,2009Natur.458..481N} and V404 Cygni \citep{2016Natur.534...75M,2017MNRAS.469.3141T}. However, the donor star in both of these systems is a low-mass star. Another well-known supercritical source is the Galactic microquasar SS433, which is a confirmed nonthermal emitter and might be a possible example of a SCWB in our Galaxy \citep[see][for an extensive review]{2004ASPRv..12....1F}. Many ultraluminous X-ray sources (ULXs) detected in nearby galaxies might also belong to this category of sources. 

In this paper, we explore the CWB scenario where one of the winds is launched by a supercritical disk around a black hole. We start by characterizing the disk model and the radiation fields it produces (Sections \ref{subsec: accretion disk} and \ref{subsec: radiation fields}). We then investigate the motion of particles under the radiation pressure in such fields (Section \ref{sect: wind of the disk}). This allows us to get reasonable estimates of the terminal velocities expected for the matter ejected in the direction of the companion star. We then proceed to study the wind interactions, shock adiabaticity, and other relevant issues for particle acceleration in Sect. \ref{sec: winds}. This is followed by estimates of energy losses for accelerated particles, particle distributions, and calculations of the nonthermal output (Sect. \ref{sec: radiative processes}). In Section \ref{sect: models} we present results for some specific models, with different choices of the accretor mass and the accretion power. The donor star is supposed to be a hot O.5V with a temperature of 41500 K and a kinetic power of a few times $10^{37}$ erg s$^{-1}$. We finally apply our model to the extragalactic binary system NGC 4190 ULX 1. After a discussion (Sect. \ref{sect:discussion}), we close with a summary and our conclusions.

\section{The accretion disk and its wind}

We assume that the X-ray binary is composed of a Population I star and a nonrotating stellar mass black hole (BH) in a close orbit. 

The orbital semi-axis $a$, the stellar radius, and the mass ratio of the system, $q=M_*/M_{\rm BH}$, satisfy \citep{1983ApJ...268..368E}:
\begin{equation} \label{eggleton}
    R_{\rm{lob}}^*=  \dfrac{a \ 0.49 \ q^{2/3}}{0.6 \ q^{2/3} + \ln{(1+q^{1/3})}}\mathbf{,}
\end{equation}
where $M_*$ is the mass of the star and $M_{\rm BH}$ the mass of the BH.
Hence, the star overflows its Roche lobe $R_{\rm{lob}}^*$, transfers mass to the BH through the Lagrange point, and an accretion disk is formed due to the angular momentum of the system. 

In this section, we describe the semi-analytical models we use to study the accretion disk, the spatial distribution of the radiation fields produced by the disk, and the wind ejected from its surface. We assume a Newtonian potential for the gravity field, because we are interested in weak-field processes.

\subsection{Accretion disk} \label{subsec: accretion disk}

We adopt cylindrical coordinates with axial symmetry along the $z$-axis, neglect the self-gravity of the disk gas, and consider a nonmagnetized disk with a super-Eddington accretion rate at the outer part of the disk, $\dot{m}_{\rm input}=\dot{M}_{\rm input}/\dot{M}_{\rm Edd} \gg 1$, where $\dot{M}_{\rm input}$ is the input of mass per time unit in the accretion disk. 
The Eddington rate is given by
\begin{equation} \label{tasa critica}
    \Dot{M}_{\rm{Edd}}=   \frac{L_{\rm{Edd}}}{\eta c^2} \approx 2.2\times 10^{-8} M_{\rm BH} \ {\rm yr^{-1}} = 1.4 \times 10^{18} \frac{M_{\rm BH}}{M_\odot} \ \rm{g \, s^{-1}},
\end{equation}
with $L_{\rm Edd}$ the Eddington luminosity\footnote{The Eddington luminosity is defined as the luminosity required to balance the attractive gravitational pull of the accreting object by radiation pressure.}, $\eta \approx 0.1$ the accretion efficiency, and $c$ the speed of light.

The critical or spherization radius, given by
\begin{equation}
    r_{\rm crit} \sim  40 \dot{m}_{\rm input} r_{\rm g},\label{eq:rg}
\end{equation}
separates the disk in two regions: a standard outer disk \citep{1973A&A....24..337S} and a radiation-dominated inner disk with advection \citep{2004PASJ...56..569F}. In relation (\ref{eq:rg}), $r_{\rm g}=GM_{\rm BH}/c^2$ is the gravitational radius of the BH, with $G$ the gravitational constant.
In the disk model, the advection is parameterized as a fraction $f$ of the viscous heating, $Q_{\rm adv}=fQ_{\rm vis}$, and the disk becomes geometrically thick in the inner region, where the ejection of winds by the radiation force helps to regulate the mass-accretion rate onto the BH ($\dot{M}_{\rm acc}$) at the Eddington rate\footnote{$\dot{M}_{\rm acc}=\dot{M}_{\rm input}$ in the outer region of the disk and $\dot{M}_{\rm acc}=\dot{M}_{\rm input}r_{\rm d}/r_{\rm crit}$ in the inner region \citep{2004PASJ...56..569F}.}. 

As the disk is optically thick, we assume that it radiates locally as a blackbody. The radiation intensity  of a plasma element in the comoving frame of the outer and inner disk, at a radius $r_{\rm d}$ measured on the equatorial plane, is

\begin{equation}\label{intensidad}
     I_0=\frac{1}{\pi}\sigma T_{\rm eff}^4 =
    \left\lbrace \begin{array}{l}
    \dfrac{1}{\pi}\dfrac{3GM_{\rm BH}\dot{M}_{\rm input}}{8\pi r_{\rm d}^3} f_{\rm in}, \ \ r_{\rm d} > r_{\rm crit}\\ \\
  \dfrac{1}{\pi}\dfrac{3}{4}\sqrt{c_3}\dfrac{L_{\rm Edd}}{4\pi r_{\rm d}^2}, \ \ r_{\rm d} \le r_{\rm crit},
    \end{array} 
    \right.
\end{equation}
where $\sqrt{c_3}=H/r_{\rm d}=\tan{\delta}$, with $H$ the scale height of the disk, $\delta$ the disk opening angle, and $f_{\rm in}=1-r_{\rm in}/r_{\rm d}\approx1$ \ (as $r_{\rm d}>r_{\rm crit}$, then $r_{\rm d}\gg r_{\rm in})$. Here, $c_3$ (along with $c_1$ and $c_2$ used in the following section) is a coefficient that depends on the advection parameter, the adiabatic index of the gas $\gamma$, and the viscosity $\alpha$ \citep[see Appendix in][]{2004PASJ...56..569F}. We adopt a disk with $f=0.5$ and $\alpha=0.5$; that is, we assume equipartition between advection and viscous heating. The index $\gamma=4/3$ corresponds to a radiation-dominated gas in the inner disk. These values lead to a disk-opening angle of  $\delta=30^{\circ}$.

\subsection{Radiation fields} \label{subsec: radiation fields}
The wind launched from the radiation-dominated region of the disk will be determined by the radiation forces acting upon the particles on the disk surface and along their subsequent trajectories. These forces will have contributions from different parts of the disk in relative motion with respect to the particles. Some radiation will be blueshifted and some will be redshifted, resulting in differential azimuthal forces onto the particles and then transferring angular momentum from the disk to the wind. 

\begin{figure}
    \centering
\includegraphics[width=\columnwidth]{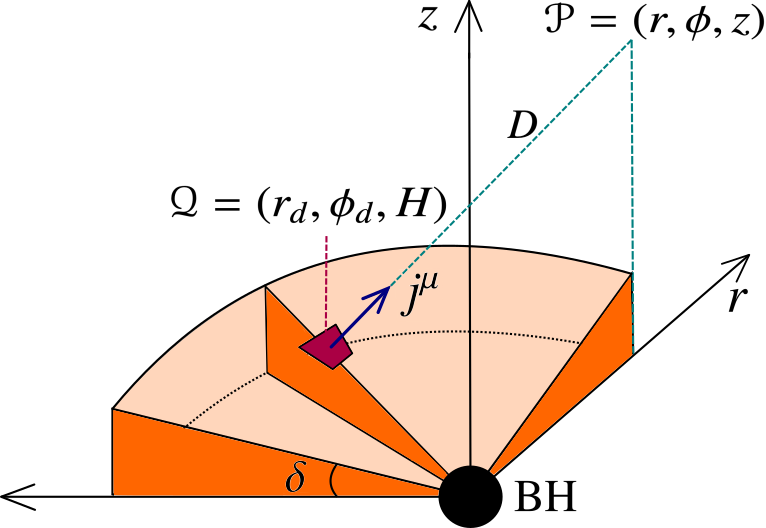}
    \caption{Geometry of the present disk model. The radiation fields are calculated in the $rz$ plane, where $\phi=0$. Here, $\mathscr{Q}$ is the position of the plasma element of the disk and $\mathscr{P}$ the point of calculation on the $rz$ plane. The scale height of the disk is $H$, and $D$ is the distance between $\mathscr{Q}$ and $\mathscr{P}$. The short arrow is the direction cosine $j^{\mu}$. This figure is adapted from \cite{1999PASJ...51..725W}.}
    \label{fig:disk}
\end{figure}

In order to obtain the radiative contribution of each plasma element  $\mathscr{Q}=(r_d,\phi_d,H) $ of the disk surface, at any point $\mathscr{P}=(r,\phi,z)$ above or below the disk, we make a transformation of the intensity between the inertial and comoving reference frames (see Fig. \ref{fig:disk}). Azimuthal symmetry allows us to perform the calculations for any constant value of $\phi$; therefore, we do it in the $rz$ plane $(\phi=0)$.
The relativistic Doppler factor $\mathscr{D}$ provides the transformation between the reference frames \citep{1980AmJPh..48..612M}:
\begin{equation} \label{intensidad}
    I=\mathscr{D}^4 I_0=\frac{I_0}{(1+z_{\rm red})^4},
\end{equation}
where $z_{\rm red}$ is the redshift factor given by \citep{1999PASJ...51..725W}
\begin{equation}
    z_{\rm red}=-\frac{(r \cos{\phi_{\rm d}} - r_{\rm d})v_r - (r \sin{\phi_{\rm d}}) v_\phi + (z-H)v_r c_3}{cD}.
\end{equation}
Here, $D$ is the distance between $\mathscr{P}$ and $\mathscr{Q}$, $v_\phi=c_2 v_{\rm K}$ is the azimuthal velocity and $v_r=-c_1 \alpha v_{\rm K}$ is the radial velocity, with $v_{\rm K}=\sqrt{GM_{\rm BH}/r_{\rm d}}$ the Keplerian velocity. We note that we only consider the inner part of the disk for these calculations, because the intensity decays with $r_{\rm d}^{-3}$.

The radiation-field tensor is given by \citep{Rybicki1986}
\begin{equation} \label{eq: radiation tensor}
    R^{\mu \nu}=\begin{pmatrix}
    E & \frac{1}{c}F^{\alpha} \\ 
    \frac{1}{c}F^{\alpha} & P^{\alpha \beta}
    \end{pmatrix}= \frac{1}{c}\int Ij^{\mu}j^{\nu} \rm{d}\Omega.
\end{equation}
This is a symmetric tensor of rank $2$ and therefore we calculate ten elements in total: one for the energy density $E$, three for the flux vector $F^{\alpha}$, and six for the stress tensor $P^{\alpha \beta}$. In Eq. \ref{eq: radiation tensor}, $j^{\mu}$ and $j^{\nu}$ are the direction cosines in Cartesian coordinates, and $\Omega$ is the solid angle subtended by $\mathscr{Q}$:
\begin{equation}
    j^\mu=\left(\frac{r-r_{\rm d} \cos{\phi_{\rm d}}}{D},\frac{-r_{\rm d} \sin{\phi_{\rm d}}}{D},\frac{z-H}{D}\right),
\end{equation}
\begin{equation}
    {\rm{d}} \Omega = \frac{-(r\cos{\phi_{\rm d}}-r_{\rm d})\sin{\delta}+(z-H)\cos{\delta}}{D^3} \,{\rm d}S,
\end{equation}
where ${\rm d}S=\sqrt{1+c_3}\,r_{\rm d}\,{\rm{d}} r_{\rm d} \,{\rm{d}} \phi_{\rm d}$.

\subsection{Particles in the photon field} \label{sect: wind of the disk}

We now calculate the trajectory and velocity of the particles ejected from the disk when they interact with photons of the ambient radiation field.

The equation of motion under a relativistic, radiation treatment, is given by \citep{2020fafd.book.....K}
\begin{equation}\label{conservation}
    f_{\mu} = -\frac{\partial \Phi_{\rm e}}{\partial x^{\nu}} + R_{\mu;\nu}^{\nu},
\end{equation}
where $f_{\mu}$ is the four-force per unit volume. The effective potential $\Phi_{\rm e}$ is the sum of gravitational $(\Phi_{\rm g})$ and centrifugal $(\Phi_{\rm c})$ potentials. The semicolon $(;)$ in the second term refers to the covariant differentiation of the energy-momentum tensor. 

As we consider a disk with axial symmetry, the gravitational potential cancels out in the azimuthal coordinate: $\partial \Phi_{\rm g}/\partial x^{\alpha}=(\partial \Phi_{\rm g}/\partial r,0,\partial \Phi_{\rm g}/\partial z)$.
Furthermore, the centrifugal potential acts only in the radial direction: $\partial \Phi_{\rm c}/\partial x^{\alpha}=(l^2/r^3,0,0)$, with $l=r_{\rm d}^2\omega_{\rm K}$ being the specific angular momentum of the disk, and  $\omega_{\rm K}$ the angular velocity.

The equations of motion of the ejected particles can be found working with Eq. \ref{conservation}. In terms of the nondimensional form of the  radiation-field tensor elements $\epsilon$, $f^{\alpha}$, and $p^{\alpha\beta}$,
the system of differential, tensorial, and coupled equations is as follows \cite[equations originally derived by][Eq. 42--44, but now extended to second order in velocity]{1999PASJ...51..725W}:

\noindent{Radial coordinate:}
\begin{flalign} \label{eq: radial}
 \frac{{\rm d}u^r}{{\rm d}\tau}=&  -\frac{\partial \Phi_{\rm g}}{\partial r} + \frac{l^2}{r^3} + &&\\ \nonumber &+\frac{1}{2}[\gamma f^{r} - p^{r \beta}u_{\beta} - \gamma^2 \epsilon u^{r} + u^{r}(2\gamma f^{\beta} u_{\beta} - p^{\beta \delta}u_{\beta} u_{\delta})].&&
\end{flalign}

\noindent{Azimuthal coordinate:}
\begin{flalign} \label{eq: azimuthal}
\frac{1}{r}\frac{{\rm{d}}l}{{\rm d}\tau}  =& \ \frac{1}{2}[\gamma f^{\phi} - p^{\phi \beta}u_{\beta} - \gamma^2 \epsilon (l/r) + &&\\\nonumber 
 &+(l/r)(2\gamma f^{\beta} u_{\beta} - p^{\beta \delta}u_{\beta} u_{\delta})].&&
 \end{flalign}

\noindent {Height coordinate:}
\begin{flalign} \label{eq: height}
 \frac{{\rm d}u^z}{{\rm d}\tau}=&
 -\frac{\partial \Phi_{\rm g}}{\partial z} + &&\\ \nonumber
 &+ \frac{1}{2}[\gamma f^{z} - p^{z \beta}u_{\beta} - \gamma^2 \epsilon u^{z} + u^{z}(2\gamma f^{\beta} u_{\beta} - p^{\beta \delta}u_{\beta} u_{\delta} )],&&
\end{flalign}
where $u^{\mu}$ denotes the four-velocity of the particles and $\gamma$ the Lorentz factor, which is given by
\begin{equation} \label{lorentz}
    \gamma=\sqrt{1+u^ru^r+l^2/r^2+u^zu^z}.
\end{equation}
The free parameter of these equations of motion is the launching radius of the particles, $r_0$, and we assume as initial condition that the particles co-rotate with the disk at this radius, $u_0^{\alpha}=(0,l_0/r_0,0)$. 

We solve this system of equations  numerically and assume that the kinematics of the disk-driven wind is roughly described by the trajectory and terminal velocities obtained for the test particles. As the accretion rate in the inner region of the disk is regulated at the Eddington rate, the mass loss in the wind is of the order of the super-Eddington accretion rate, $\dot{M}_{\rm dw}\sim \dot{M}_{\rm input}$.

\section{Collision of winds} \label{sec: winds}

The wind ejected from the disk collides with the stellar wind at the interaction region, where shocks are generated giving rise to particle acceleration. An important quantity that characterizes the wind is the kinetic luminosity, $L_{\rm K}=\dot{M}v^2/2$, where $\dot{M}$ is the mass-loss rate and $v$ the velocity of the fluid. A small fraction of the total kinetic power of the wind is transferred to relativistic particles, $L_{\rm rel}\sim 0.1L_{\rm K}$, where we assume equipartition between relativistic protons and electrons ($L_{\rm e}=L_{\rm p}$). The mass-loss rate and velocity of the stellar wind are set according to the parameters found in the literature for the type of star we have chosen \citep[e.g.,][]{2019AJ....158...73K}. 
In the case of the disk-driven wind, the velocity is obtained following the procedures described in the previous section. Given the orbital separation, the disk inclination, and the stellar size, we estimate that $\sim 10\%$ of the original kinetic power 
reaches the acceleration region. We assume a circular orbit, that is, the geometry associated with the collision of winds does not depend on the orbital phase.

In this section, we describe the models for the collision region, the magnetic ambient field, and the shocks. We adopt a one-zone approximation for these calculations.

\begin{figure}[h] 
        \centering
            \includegraphics[width=\columnwidth]{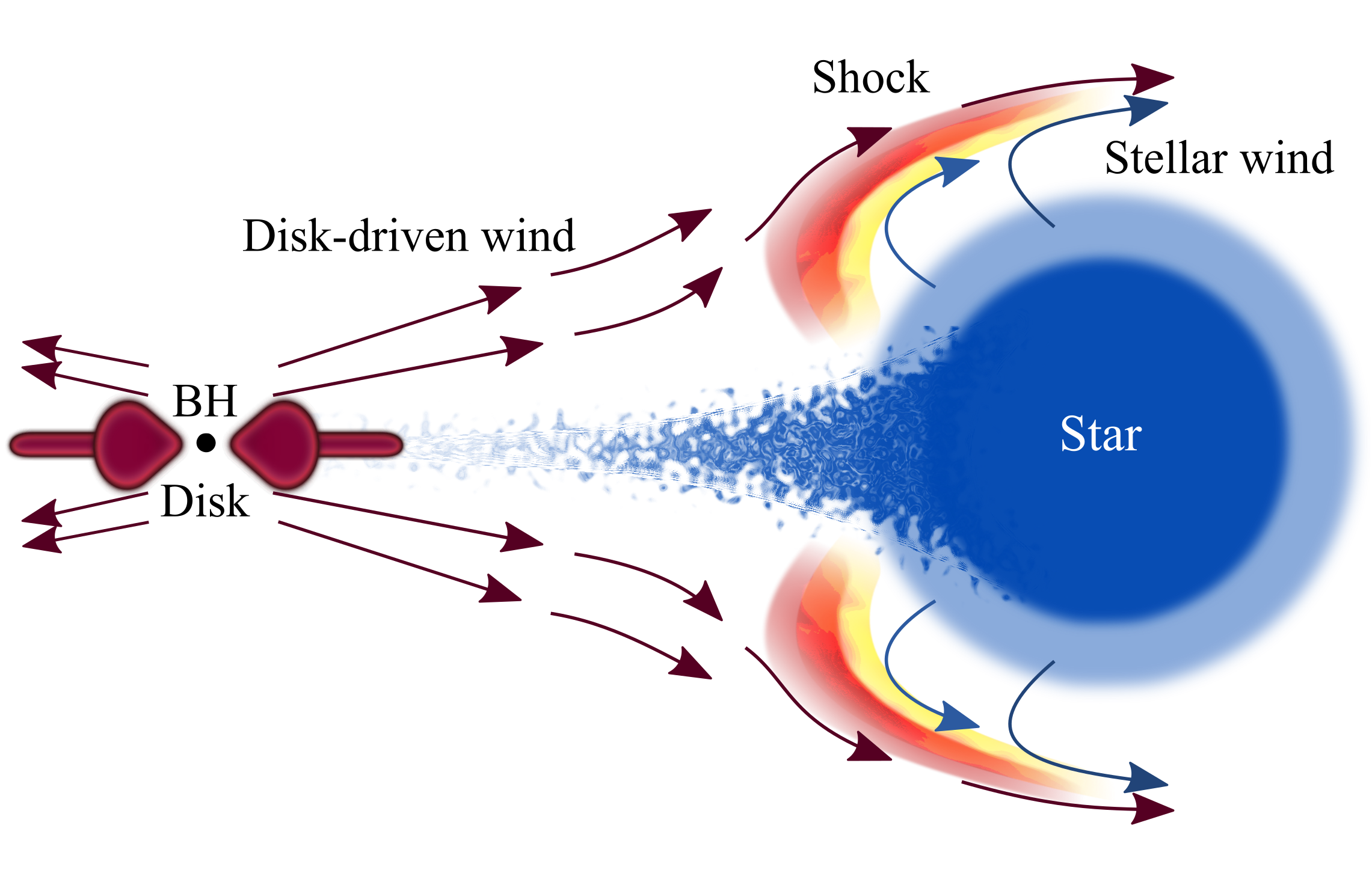}
            \caption{Scheme of the wind collision seen in the $rz$ plane (not to scale), adapted from \cite{2021BAAA...62..262A}.}
\label{fig: collision}
\end{figure}

\subsection{Contact discontinuity} \label{sect: cd}

The winds collide at a surface called the contact discontinuity (CD).  
The stagnation point (SP) is the closest position of the CD to the star, and is located where the ram pressures of the winds are in equilibrium,
\begin{equation}\label{eq: presiones}
    P_{\rm{ram}}(r_{\rm{BH}})=\rho_{\rm{dw}} v_{\rm{dw}}^2=\rho_{\rm{*w}} v_{\rm{*w}}^2=P_{\rm{ram}}(r_{\rm{*}}).
\end{equation}
Here, $r_{\rm{BH}}$ and $r_{\rm{*}}$ are the distances to the SP from the BH and from the center of the star, respectively. The density of the spherical stellar wind at this location is given by
\begin{equation} \label{densidad viento estrella}
    \rho_{\rm *w}=\dfrac{\dot{M}_*}{4\pi r_{\rm{*}}^2 v_{\rm *w}},
\end{equation} 
whereas the density of the disk-driven wind reads 
\begin{equation} \label{densidad viento disco}
    \rho_{\rm{dw}}=\dfrac{\dot{M}_{\rm{dw}}}{\Omega r_{\rm{BH}}^2 v_{\rm{dw}}},
\end{equation}
where $\Omega = 2\pi (1- \cos \theta)$ is the solid angle of the wind and $\theta$ the semi-opening angle of the wind. 
Solving these equations we obtain the position of the SP. 

\subsection{Magnetic field}

The strength of the magnetic field at the CD is essentially determined by the stellar surface magnetic field $B_*$.
The intensity of $B_{\rm CD}$ and its topology --dipole (i), radial (ii), or toroidal (iii)--, is given by \citep{1993ApJ...402..271E}:
\begin{equation} \label{magnetic_field}
    B_{{\rm CD}} \approx B_* 
    \times \left\lbrace \begin{array}{l}
    R_*^3/r_{\rm *}^3, \ \  R_*<r_{\rm *}<r_{\rm A}, \ \hfill ({\rm i}) \\ \\
    R_*^3/r_{\rm A}r_{\rm *}^2, \ \ r_{\rm A}<r_{\rm *}<R_*(v_{\rm *w}/v_*^{\rm rot}), \ \hfill ({\rm ii}) \\ \\
    R_{*}^2v_*^{\rm rot}/r_{\rm A}r_{\rm *}v_{\rm *w}, \ \ R_*(v_{\rm *w}/v_*^{\rm rot})<r_{\rm *}, \ \  \hfill ({\rm iii}),
    \end{array}
    \right.
\end{equation}
where $R_*$ is the stellar radius, $r_{\rm A}$ the Alfv\'en radius, and $v^{\rm rot}_* \sim 0.1 v_{\rm *w}$ the surface rotation velocity.

\subsection{Particle acceleration and shock} \label{sect: acceleration}

Particles are accelerated up to relativistic energies in the collision region through a first-order diffusive shock mechanism.
Two shock fronts are generated: a forward shock (FS) that propagates through the stellar wind, and a reverse shock (RS) that propagates through the  wind of the disk. The diffusive acceleration rate of the particles is given by
\citep[e.g.,][]{1999tcra.conf..247P}:
\begin{equation}
    t^{-1}_{\rm{ac}}=\eta_{\rm ac}~\frac{e~Z~c~B_{\rm CD}}{E},
\end{equation}
where $e$ is the electric charge, $Z$ the atomic number, and $E$ is the energy of the particle. The acceleration efficiency, $\eta_{\rm ac}$, depends on the diffusion coefficient of the particles, the shock velocity, and the angle between the magnetic field and the normal to the shock plane. We assume that the shock propagates perpendicular to the magnetic field and that diffusion occurs in the Bohm regime. Thus, the acceleration efficiency is
\begin{equation} \label{eq: eficiencia}
    \eta_{\rm ac} \approx \frac{3}{8} \left(\frac{v_{\rm sh}}{c}\right)^2,
\end{equation}
where the shock velocities in the reference frame where one of the fluids is at rest, $v_{\rm *w}=0$, and the other one moves with a velocity $v_{\rm dw}$, are given by \citep{1996ApJ...464..131L}:
\begin{equation}\label{eqn:reverse}
    v_{\mathrm{RS}}=-\frac{4}{3}\frac{1}{1+\sqrt{n_{\mathrm{*w}}/n_{\mathrm{dw}}}}v_{\mathrm{dw}},
\end{equation}
\begin{equation}\label{eqn:reverse}
    v_{\mathrm{FS}}=\frac{4}{3}\frac{1}{1+\sqrt{n_{\mathrm{dw}}/n_{\mathrm{*w}}}}v_{\mathrm{dw}}.
\end{equation}
Here, $n_{\rm *w}$ and $n_{\rm dw}$ are the numerical densities of the winds ($n_{\rm w}=\rho_{\rm w}/m_{\rm p}$, with $m_{\rm p}$ the mass of the proton). The pressure and density of the shocked medium are calculated following the Rankine-Hugoniot relations \cite[e.g.,][]{1999isw..book.....L}.

As we are interested in the nonthermal particle distribution, we investigate only adiabatic shocks; that is, where radiative losses are negligible. This is because in radiative shocks the gas in the shocked region emits large amounts of thermal radiation; the system therefore loses energy, the entropy increases, and the medium becomes increasingly homogeneous. If magnetic-inhomogeneities disappear, the acceleration efficiency decays abruptly, aborting the formation of nonthermal distributions. 

The shock is adiabatic if the thermal cooling length $R_{\mathrm{\Lambda}}$ is larger than the size of the acceleration region $\Delta x_{\rm{ac}}$ \citep{1979ARA&A..17..213M}. The cooling length reads
\begin{equation} \label{longitud enfriamiento}
    R_{\mathrm{\Lambda}} = \dfrac{5.9 \times 10^{11} \mu (v_{\rm{sh}}/\rm{km \ s^{-1}})^3}{(n_{\rm{w}}/{\rm{cm^{-3}}})[\Lambda (T_{\rm sh}) / \rm{erg \ s^{-1} \ cm^{-3}}]} \ \rm{cm}.
\end{equation}
Here, $n_{\rm w}$ is the number density of the undisturbed medium, $\mu$ is the average molecular weight ($\mu=0.6$ for a fully ionized plasma), and $\Lambda(T_{\rm sh})$ is the cooling function, which depends on the shock temperature \citep{1976ApJ...204..290R,1998MNRAS.298.1021M,2003ApJ...587..278W}. This latter function can be written as
\begin{equation}
    \Lambda(T_{\rm sh}) = \left\{ \begin{array}{ll}
    4\times 10^{-29} T_{\rm sh}^{0.8},  & \ 55 \ {\rm{K}} \le T_{\rm sh} < 10^4 \ {\rm{K}} \\
    7\times 10^{-27} T_{\rm sh},  & \ 10^4 \ {\rm {K}} \le T_{\rm sh} < 10^5 \ {\rm {K}} \\
    7\times 10^{-19} T_{\rm sh}^{-0.6},  & \ 10^5 \ {\rm {K}} \le T_{\rm sh} < 4\times10^7 \ {\rm {K}} \\ 
    3\times 10^{-27} T_{\rm sh}^{0.5},  & \  T_{\rm sh} \ge 4\times 10^7 \ {\rm {K,}}
    \end{array}
    \right.
\end{equation}
where $T_{\rm sh}$ is given by
\begin{equation}
    T_{\rm sh}=18.21 \mu \left(\frac{v_{\rm sh}}{\rm km \ s^{-1}}\right)^2 \rm K.
\end{equation}
We note that this temperature has a maximum value in a collisional plasma: it is self-regulated by the pair-creation, satisfying in any case $k_{\rm B}T_{\rm sh}<1\ \rm MeV$ ($k_{\rm B}$ is the Boltzmann constant). 

We assume that the size of the acceleration region is a fraction of the distance from the BH to the SP, $\Delta x_{\rm{ac}}\sim 0.1r_{\rm BH}$. As we consider a one-zone model, the acceleration region must be narrow enough to generate near-homogeneous conditions.

\section{Radiative processes} \label{sec: radiative processes}

Particles accelerated at the shock can cool through different processes and produce nonthermal radiation. The timescales associated to this cooling are related to the total energy-loss of the particles:
\begin{equation}
    \frac{dE}{dt}\approx \frac{-E}{t_{\rm{cool}}},
\end{equation}
where the total cooling rate is
\begin{equation}
    t_{\rm{cool}}^{-1} = \sum_i t_{i}^{-1},
\end{equation}
where $t_i$ corresponds to each timescale of the involved cooling processes.

We assume advective escape; that is, particles are removed from the acceleration region by the bulk motion of the fluid. If the timescales of cooling are shorter than those of escape, particles radiate before they escape from the acceleration region. The maximum energy for each kind of particle can be inferred by looking at the point where the acceleration rate is equal to the total cooling or escape rate. This energy cannot exceed the maximum energy imposed by the Hillas criterion, $E^{\rm max}_{\rm e,p}<E^{\rm max}_{\rm Hillas}$.

As we are interested in nonthermal processes, we work at scales smaller than the size of the binary system and assume that rotation effects are negligible there. Effects caused by the orbital motion, such as Coriolis or centrifugal forces, could be relevant on larger scales and lead to strong disturbances in the flow and thermal processes. The analysis of such effects usually requires numerical simulations and is beyond the scope of this work.

\subsection{Energy losses}
We consider adiabatic and radiative losses. Adiabatic cooling is related to the work done by the particles of the wind to expand the shocked gas. Radiative cooling is caused by nonthermal processes as a consequence of the interaction of the wind particles with ambient fields and matter.

Our model is lepto-hadronic, and so we calculate the following radiative processes  numerically:

--Synchrotron: interaction of protons and electrons with the ambient magnetic field, which will be amplified by a factor of 4 in the shocked region due to Rankine-Hugoniot relations. 

--Inverse Compton (IC): collision of relativistic electrons with photons of the ambient radiation field.

--Bremmstrahlung: Coulombian interactions between relativistic electrons and cold matter.

--Photo-hadronic interactions: interaction of highly relativistic protons with photons of the ambient radiation field.

--Proton-proton: collision of relativistic protons with cold matter.

In addition, we take into account inelastic collision of particles with atoms of the dense medium; that is, ionization losses, which can be relevant in the 1--100 MeV range. We note that in this energy range, ionization losses largely dominate over Coulomb scatterings \citep[see e.g., Fig. 7 from][]{1996A&A...309.1002O}, and so the latter are not included in our analysis. The reader is referred to \cite{Romero&Paredes2011}, \cite{Romero-Vila2014}, and \cite{2020A&A...636A..92M} plus references therein for additional details on radiative processes.

\subsection{Particle distribution} \label{subsec: particle distribution}

We investigate the evolution of particles that are accelerated at the shock and injected into the surrounding medium. The medium around the shock is the shocked gas of the winds. In this paper, we restrict our analysis to this region. Beyond the binary, the surrounding medium has been affected by the effects of the stellar winds, and so the system is expected to be located inside a bubble inflated by the winds and surrounded by a shell formed with the swept-up material at distances of a few to several  parsecs, depending on the mass of the black hole progenitor.  Inside the bubble, where the advected protons will be injected, the density is expected to be lower than that of the standard interstellar medium (e.g., around 0.01 cm$^{-3}$ or less). In the shell, there should be sufficient material for hadronic interactions with the protons diffused or transported from the central source\footnote{These effects will be discussed elsewhere; some of them might be responsible for part of the high-energy emission observed in the shell of W50, which is powered by SS433, although there are jets involved in this specific object.}.

The relativistic particles have a distribution given by ${\rm d}N= n(\vec{r},E,t){\rm d}E{\rm d}V$, where $n$ is the number density of particles, $t$ the time, $\vec{r}$ the position, $V$ the volume, and $E$ the energy. The evolution of this distribution is determined by the transport equation \citep[see e.g.,][]{1964ocr..book.....G,Romero&Paredes2011}. We solve this equation numerically in steady state and in the one-zone approximation: 
\begin{equation}
    \frac{\partial}{\partial E}\left[\frac{{\rm d}E}{{\rm d}t} N(E)\right]+\frac{N(E)}{t_{\rm esc}}=Q(E),
\end{equation}
where $t_{\rm esc} \sim \Delta x_{\rm ac}/v_{\rm sh}$ is the advection time, and the particle injection function,
\begin{equation} \label{eq: injection} 
    Q(E)=Q_{0}E^{-p}\exp{(-E/E_{\rm max})},
\end{equation} 
is a power-law in the energy with an exponential cutoff and a spectral index $p=2.2$, which is characteristic of the Fermi first-order acceleration mechanism \citep[see e.g.,][]{1983RPPh...46..973D}. The normalization constant $Q_0$ is obtained from
\begin{equation}
    L_{(\rm e,p)}=\Delta V \int^{E^{\rm max}_{\rm (e,p)}}_{E^{\rm min}_{\rm (e,p)}}{\rm d}E_{\rm (e,p)}E_{\rm (e,p)}Q_{\rm (e,p)}(E_{\rm (e,p)}),
\end{equation}
where $\Delta V$ is the volume of the acceleration region, and $E^{\rm max}_{\rm (e,p)}$ the maximum energy reached by protons and electrons, which is found by looking at the point where the acceleration rate is equal to the total cooling or escape rate.

\subsection{Nonthermal emission}

Once we have the particle distributions, we calculate the spectral energy distribution (SED) for each of the relevant processes involved in cooling. We find that in SCWBs, electrons typically cool by synchrotron and IC mechanisms, and protons escape from the acceleration region without significant cooling. The resultant nonthermal SED usually yields a broadband spectrum from radio waves (due to synchrotron emission) to gamma-rays (due to IC emission).

\subsection{Wind emission}

We calculate the thermal emission of the photosphere of the disk-driven wind assuming a spherically symmetric wind that expands with constant velocity equal to its terminal velocity. Since the mass-loss rate of the disk is much higher than the critical rate, the wind is optically thick and therefore we assume that it radiates locally as a blackbody. The temperature measured by an observer at infinity is given by \citep{2009PASJ...61.1305F}:
\begin{equation}
     \sigma_{\rm T} T_{\rm dw}^4=\frac{\dot{e} \, L_{\rm Edd}}{(1-\beta \cos{\Theta})^4 \, 4 \pi R^2}
,\end{equation}
where $\dot{e}=\dot{E}/L_{\rm Edd}$ is the normalized comoving luminosity, $\beta=v_{\rm dw}/c$ the normalized velocity, $\Theta$ the angle of the flow with respect to the line of sight, and $R=\sqrt{r^2+z^2}$, with $r$ and $z$ the being cylindrical coordinates. We assume that the comoving luminosity is equal to the Eddington luminosity ($\dot{e}=1$), as is commonly done in supercritical wind-models \citep[e.g.,][]{2009PASJ...61.1305F}.

The apparent photosphere of this wind is defined as the surface where the optical depth $\tau_{\rm photo}$ is unity for an observer at infinity. If the velocity of the wind is relativistic, the optical depth in the observer frame depends in general on the magnitude of the velocity and the viewing angle. The location of the apparent photosphere from the equatorial plane $z_{\rm photo}$ is \citep{2009PASJ...61.1305F}:
\begin{equation}
    \tau_{\rm photo}=\int^\infty_{z_{\rm photo}} \gamma_{\rm dw}(1-\beta \cos{\Theta}) \, \kappa_{\rm co} \,\rho_{\rm co} {\rm d}z =1,
\end{equation}
where $\gamma_{\rm dw}$ is the wind Lorentz factor, $\kappa_{\rm co}$ the opacity in the comoving frame, and $\rho_{\rm co}$ the wind density in the comoving frame. As we assume a fully ionized wind, the opacity is dominated by free electron scattering ($\kappa_{\rm co}=\sigma_{\rm T}/m_{\rm p}$).

\subsection{Absorption} \label{subsec: absorption}

Finally, we calculate the gamma absorption by pair creation from photon--photon annihilation, $\gamma + \gamma \rightarrow e^+ + e^-$. The  nonthermal photons in their way out of the acceleration region can find photons of the ambient radiation fields and annihilate. The absorption is quantified by the optical depth of the medium, $\tau_{\gamma \gamma}$. 
If the original luminosity of gamma rays is $L^0_\gamma(E_\gamma)$, the attenuated luminosity reads:
\begin{equation}
    L_{\gamma}(E_\gamma)=L^0_\gamma(E_\gamma)\cdot {\rm e}^{- \tau},
\end{equation}
where ${\rm{e}}^{-\tau}$ is the attenuation factor.
The targets of the ambient radiation fields are photons from the star and from the disk-driven wind photosphere.

The process of annihilation is possible only above a kinematic energy threshold given by
\begin{equation}
    E_{\gamma}E_{\rm ph} > (m_{\rm e}c^2)^2,
\end{equation}
in a frontal collision, where $E_{\rm ph}$ is the energy of the targets. The opacity caused by a photon--photon pair production for a photon created at a distance $r$ from the center of the thermal source can be obtained from \citep{Romero-Vila2008}:
\begin{equation}
    \tau_{\gamma \gamma}(E_{\gamma},r)= \int_{E_{\rm min}}^{\infty}\int_{r}^{\infty} n_{\rm ph}(E_{\rm ph},r') \, \sigma_{\gamma \gamma}(E_{\rm ph},E_{\gamma}) \, {\rm d}r' {\rm d}E_{\rm ph}, 
\end{equation}
where $n_{\rm ph}$ is the density of the ambient radiation field. The total cross-section is given by \citep[see e.g.,][]{1985Ap&SS.115..201A}:
\begin{equation}
    \sigma_{\gamma \gamma}=\frac{\pi r_{\rm e}^2}{2}(1-\xi^2)\left[(3-\xi^4)\ln{\left(\frac{1+\xi}{1-\xi}\right)}+2\xi(\xi^2-2) \right],
\end{equation}
where $r_{\rm e}$ is the classical radius of the electron, and
\begin{equation}
    \xi=\left(1-\frac{(m_{\rm e}c^2)^2}{E_{\gamma}E_{\rm ph}}\right)^{1/2}.
\end{equation}
The blackbody density radiation of the star and the photosphere of the disk-driven wind is given by
\begin{equation}
    n_{\rm ph}=\frac{2E_{\rm ph}^2}{h^3c^3} \, \frac{1}{\exp({E_{\rm ph}/k_{\rm B}T})-1},
\end{equation}
where $T$ is the temperature of the thermal source considered for each case; that is, $T_{\rm dw}$ or $T_{\rm eff}$.

On the other side, free-free absorption (FFA) must also be taken into account. The collision of low-energy photons with particles of the dense medium leads to a cutoff in the SED at radio frequencies. The denser the medium, the higher the energy at which the cutoff occurs. Therefore, FFA will determine the turnover of the synchrotron spectrum in SCWBs, which is expected to be at $\sim$GHz frequencies \citep[see e.g.,][]{Rybicki1986,2016A&A...591A.139D}.

Other absorption processes, such as the photoelectric effect, direct Compton, or $\gamma$-nucleon pair creation, are not taken into account in this paper. Their cross-sections are not high enough to become relevant in the calculation of opacity given the ambient densities that we consider here \citep[see Fig. 1 from][]{2011A&A...531A..30R}.

\section{Results} \label{sect: models} 

In this section, we apply our model to a generic super-Eddington X-ray binary. We consider a star of spectral type O.5V (Table \ref{tab:estrella}) and investigate four scenarios: in scenarios S1 and S2 we regard a BH with mass $M_{\rm BH}=5 M_{\odot}$ and mass-accretion rates of $10^2 \dot{M}_{\rm Edd}$ and $10^3 \dot{M}_{\rm Edd}$, respectively; in scenarios S3 and S4 we consider a BH with mass $M_{\rm BH}=20 M_{\odot}$ and again accretion rates of $10^2 \dot{M}_{\rm Edd}$ and  $10^3 \dot{M}_{\rm Edd}$, respectively. The complete set of parameters is summarized in Table \ref{tab: parametros resultados}.

\begin{table}[h] 
\begin{center}
\begin{tabular}{l c c}
\hline
\hline
\rule{0pt}{2.5ex}& Type O.5V Star &  \\
\hline
\rule{0pt}{2.5ex}Parameter & Value & Units \\  
\hline
\rule{0pt}{2.5ex}$M_*$    & 37 & $M_{\odot}$ \\
$R_*$   & 11  & $R_{\odot}$ \\
$T_{\rm eff}$    & $41500$ & K  \\
$\dot{M}_*$   & $1.2\times10^{-5}$ & $M_{\odot}\ \rm yr^{-1}$\\
$v_{\rm *w}$ & $2.9\times 10^{8}$ & cm $\rm s^{-1}$ \\
$v_*^{\rm rot}$ & $2.9\times 10^{7}$ & cm $\rm s^{-1}$ \\
$L_K^*$ & $3.2\times 10^{37}$ & erg $\rm s^{-1}$ \\
$B_*$ & 750 & G \\
\hline
\end{tabular}
\end{center}
\caption{Parameters adopted in the model for the star of type O.5V. All parameters from \cite{2019AJ....158...73K}, with the exception for the magnetic field \cite[from][]{2015ASPC..494...30W}.} 
\label{tab:estrella}
\end{table}

\begin{table*}[t]
\begin{center}
\resizebox{\textwidth}{!}{
\begin{tabular}{p{6cm} p{3cm} c c c c}
\hline
\hline
\multicolumn{2}{c}{} & \multicolumn{4}{c}{Scenario} \\
\cmidrule(rl){3-6}
Parameter & Symbol \ [units] & S1 & S2 & S3 & S4  \\
\hline
\rule{0pt}{2.5ex}Black hole mass$^{(1)}$ & $M_{\rm BH}$ \ [$M_{\odot}$]  & 5 & 5 & 20 & 20  \\
Mass accretion rate$^{(1)}$ & $\dot{M}_{\rm input}$ \ [$M_{\odot} \ \rm{yr}^{-1}$] & $1.1\times 10^{-5}$ & $1.1\times 10^{-4}$ & $4.4\times 10^{-5}$ & $4.4\times 10^{-4}$  \\
\hline
\rule{0pt}{2.5ex}Orbital semi-axis$^{(1)}$ & $a$ \ [$R_{\odot}$] & 15 & 15 & 22 & 22 \\
Gravitational radius$^{(2)}$ & $r_{\rm g}$ \ [$\rm{cm}$] & $7.4\times10^5$ & $7.4\times10^5$ & $2.9\times10^6$ & $2.9\times10^6$ \\
Critical radius$^{(2)}$ & $r_{\rm crit}$ \ [$r_{\rm g}$] & 4000 & 40000 & 4000 & 40000  \\
Mass loss in disk winds$^{(1)}$ & $\dot{M}_{\rm dw}$ \ [$M_{\odot} \ \rm{yr}^{-1}$] & $10^{-5}$ & $10^{-4}$ & $4.3\times10^{-5}$ & $4.3\times 10^{-4}$  \\
Kinetic power of the disk-driven wind$^{(2)}$ & $L^{\rm dw}_{\rm K}$ \ [${\rm erg \ s^{-1}}$] & $7.8\times10^{39}$ &  $7.8\times10^{40}$ &  $3.4\times10^{40}$ &  $3.4\times10^{41}$ \\
Cold matter density at SP$^{(2)}$ & $n_{\rm dw}$ \ [${\rm cm^{-3}}$] & $5.1\times10^{12}$ & $5.1\times10^{13}$ & $2.9\times10^{12}$ & $2.9\times10^{13}$  \\
Distance to SP from BH$^{(2)}$ & $r_{\rm BH}$ \ [${\rm cm}$] & $2.7\times10^{11}$ & $2.7\times10^{11}$ & $7.6\times10^{11}$ & $7.6\times10^{11}$ \\
Size of acceleration region$^{(1)}$ & $\Delta x_{\rm ac}$ \ [${\rm cm}$] & $2.7\times10^{10}$ & $2.7\times10^{10}$ & $7.6\times10^{10}$ & $7.6\times10^{10}$ \\
Shock cold matter density$^{(2)}$ & $n_{\rm RS}$ \ [${\rm cm^{-3}}$] & $2\times10^{13}$ & $2\times10^{14}$ & $1.2\times10^{13}$ & $1.2\times10^{14}$  \\
Shock cooling length$^{(2)}$ & $R_{\rm \Lambda}$ \ [${\rm cm}$] & $7.6\times10^{11}$ & $7.6\times10^{10}$ & $1.3\times10^{12}$ & $1.3\times10^{11}$  \\
Maximum energy of electrons$^{(2)}$ & $E_{\rm e}^{\rm max}$ \ [${\rm eV}$] & $10^{11}$ & $1.6\times10^{11}$ & $10^{11}$ & $10^{11}$  \\
Maximum energy of protons$^{(2)}$ & $E_{\rm p}^{\rm max}$ \ [${\rm eV}$] & $10^{15}$ & $10^{15}$ & $3\times10^{15}$ & $3.1\times10^{15}$  \\
Emission peak (low energy)$^{(2)}$ & $L_{0.01\rm mm}$ \ [${\rm erg \ s^{-1}}$] & $3.2\times10^{33}$ &  $3.2\times10^{33}$ &  $8\times10^{34}$ &  $8\times10^{34}$ \\
Emission peak (high energy)$^{(2)}$  & $L_{10\rm MeV}$ \ [${\rm erg \ s^{-1}}$] & $4\times10^{32}$ &  $4\times10^{32}$ &  $10^{34}$ &  $10^{34}$ \\
\hline
\end{tabular}}
\end{center}
\caption{Parameters of the different scenarios calculated for the model. We indicate with superscript ${(1)}$ those parameters that are assumed and with ${(2)}$ those that are derived. In all models, the system is supposed to be oriented face-on to the observer, that is, the inclination of the normal to the orbital plane $i$ with respect to the line of the sight is $\sim 0^{\circ}$.}
\label{tab: parametros resultados}
\end{table*}

\subsection{Wind}

\begin{figure}[h] 
        \centering
            \includegraphics[width=\columnwidth]{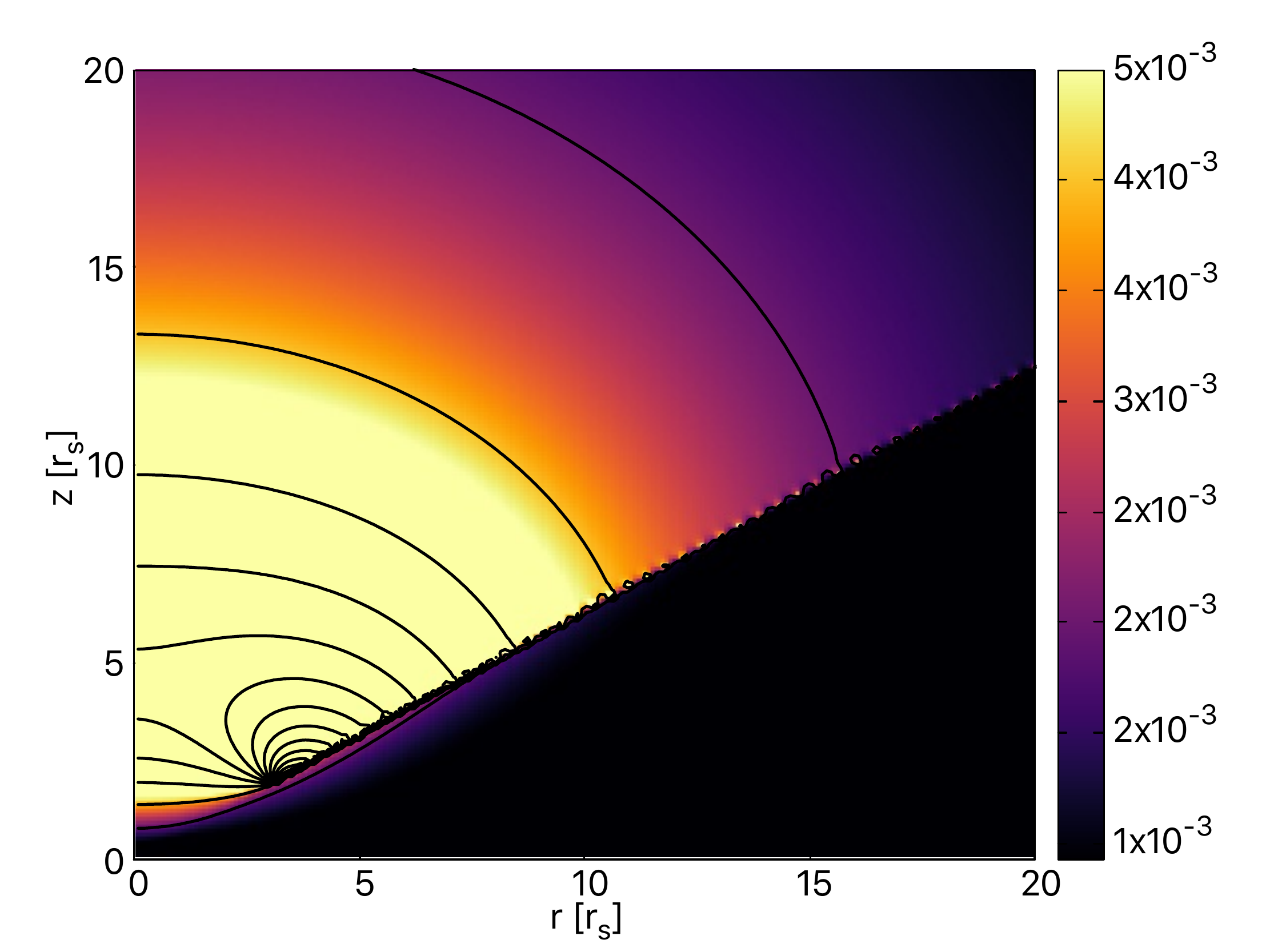}
            \caption{Contour maps of the spatial distribution of the normalized radiation energy density $\epsilon$ in the $rz$ plane above the accretion disk. Both axes are in units of Schwarzschild radius. The color bar is the intensity of $\epsilon$ and the black zone is the inflated disk ($f=0.5$, $\alpha=0.5$, $\gamma=4/3$).}
\label{fig: density}
\end{figure}

We calculate the radiation-field tensor (Eq. \ref{eq: radiation tensor}) and in Fig. \ref{fig: density} we  show the distribution of the energy density $(\epsilon)$ on the $rz$ plane, where the black zone is the inflated inner disk. We obtain a strong azimuthal flux  component of the radiation-field tensor. This distribution is the same in all four scenarios, because in the critical disk the radiation-field tensor depends on advection, viscosity, and adiabatic parameters, which remain the same in all cases. 

\begin{figure}
        \centering
            \includegraphics[width=\columnwidth]{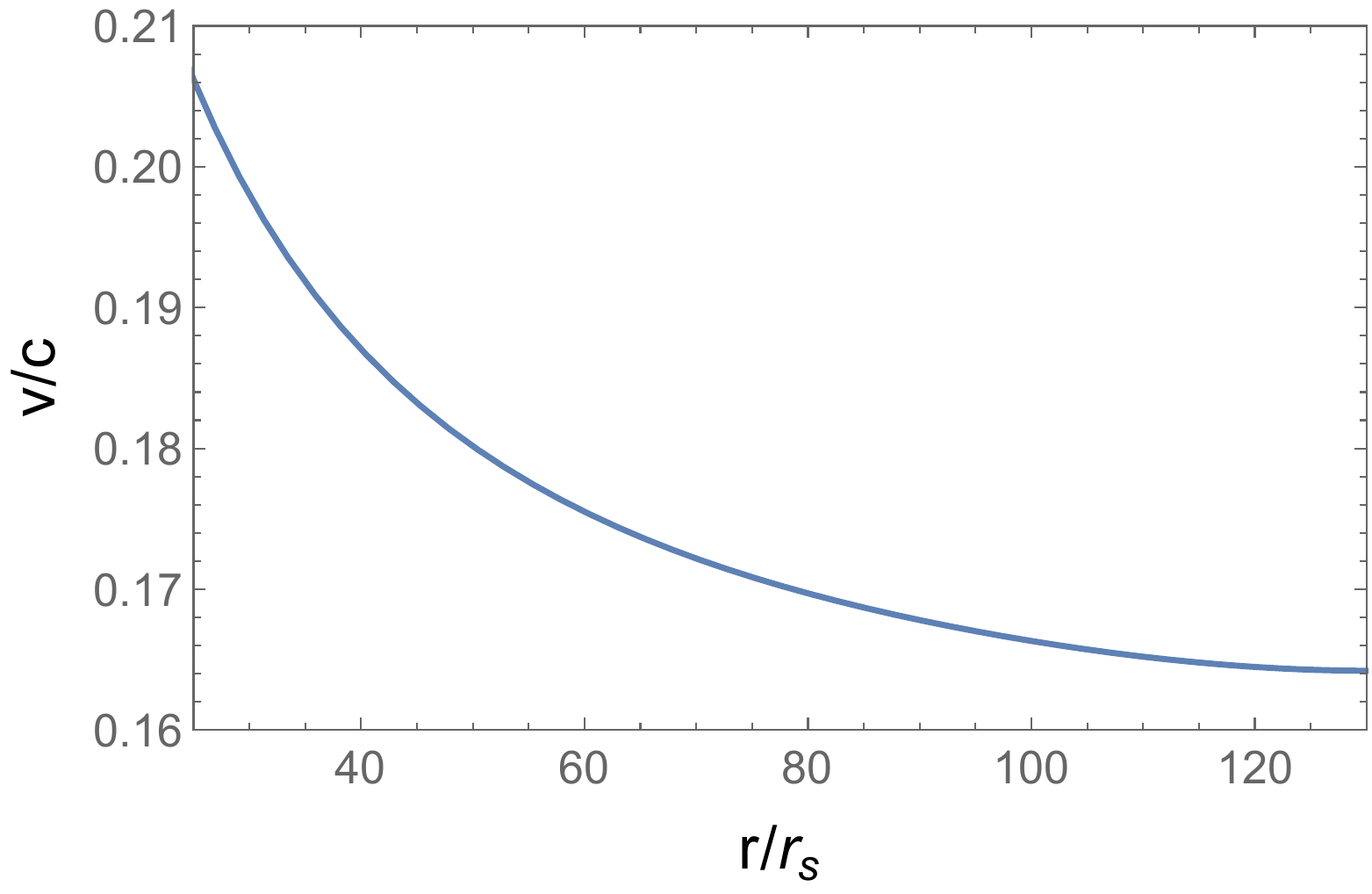}
            \caption{Normalized velocity of a wind test particle as a function of the Schwarzschild radius. The particle reaches a terminal velocity of $\sim 0.16 c$ for a launching radius of $r_0=20r_{\rm s}$ (coincident with the vertical axis).}
\label{fig: velocity}
\end{figure}

We solve Eqs. \ref{eq: radial}-\ref{eq: height} to find the trajectory and velocity of the particles. Both quantities are determined by $R^{\mu\nu}$ and therefore we obtain the same trajectories and terminal velocities in S1--S4. As an example,  in Fig. \ref{fig: velocity}  we show the normalized velocity of a test particle, with a launching radius of $40r_{\rm g}$ ($\equiv20r_{\rm s}$), which reaches a terminal velocity of $\approx0.16c$. This result does not vary much if we vary the launching radius $(\pm 0.02c$ for $\pm 20r_{\rm g})$. 

\begin{figure} 
        \centering
            \includegraphics[width=\columnwidth]{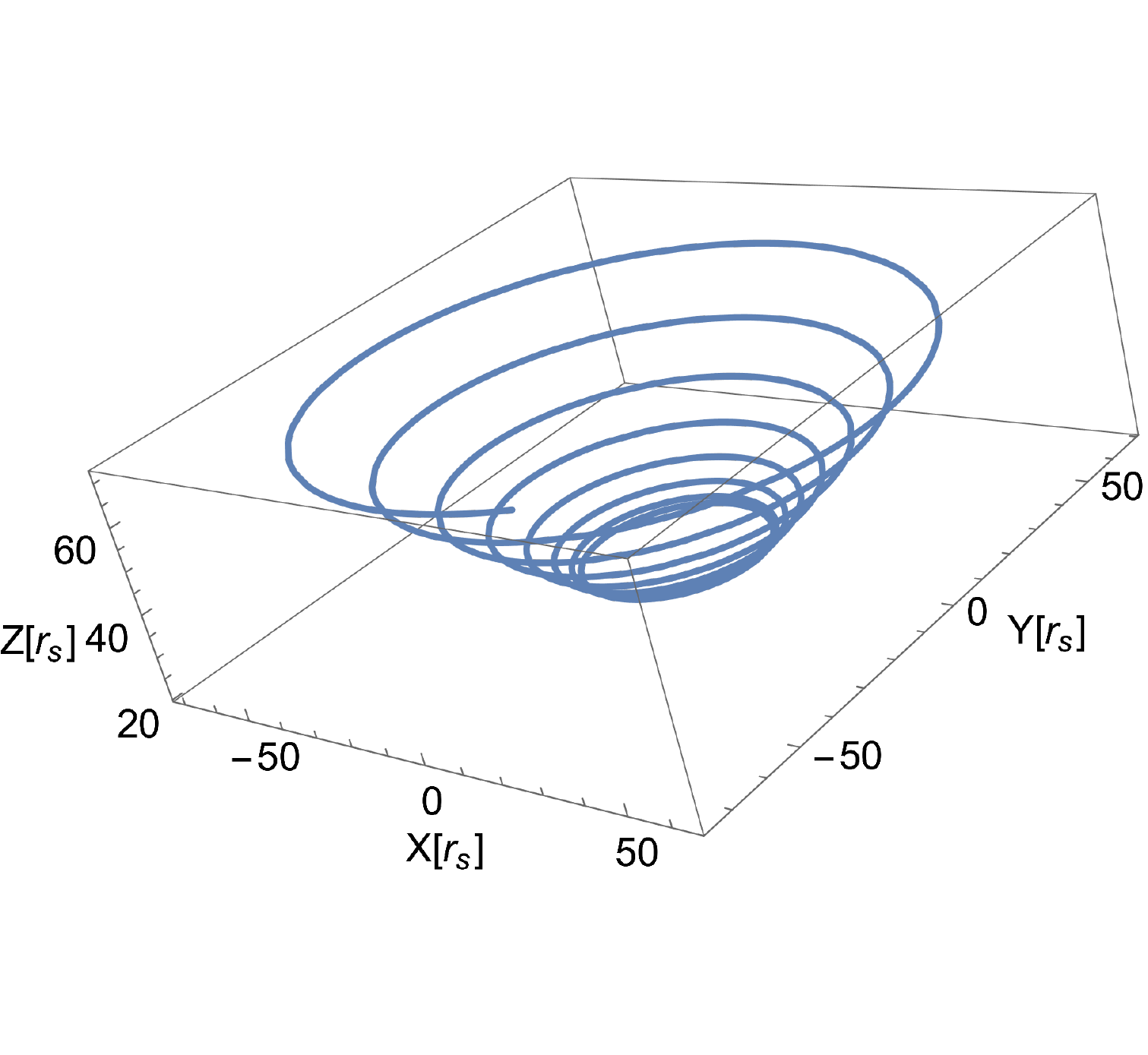}
            \caption{Trajectory of a test particle in the Cartesian 3D-space in units of Schwarzschild radius. The particles describe a helical trajectory above the inner disk because of the strong azimuthal radiation fields. The launching radius of this test particle is $r_0=20r_{\rm s}$.}
\label{fig: trajectory}
\end{figure}

The particles describe a helical trajectory in the vicinity of the BH for two main reasons (Fig. \ref{fig: trajectory}). The first is the presence of the strong azimuthal components of the radiation field, which help to maintain the spiral geometry of the particles in the inner disk. The second reason is the condition imposed for the particle ejection, namely that the particles initially have only azimuthal velocity. The intensity of the radiation field decays rapidly with distance from the BH, and therefore the ejected particles follow a spiral trajectory near the BH, but beyond a certain radius ($\sim r_{\rm crit}$) they follow a free path with a strong component of the radial velocity. 

The overall result is an equatorial wind with terminal velocities of the order of $0.15c$. The kinetic power of these winds is in the range $10^{39-41}$ erg s$^{-1}$, which is well above the power of the winds of typical WR or OB stars. Therefore, in general, the disk wind is expected to overwhelm the stellar wind.   

\subsection{Energy gain and losses}

We follow the calculations in Sect. \ref{sect: cd} and find that, in all four scenarios, the SP is located near the stellar surface and the wind of the disk   completely sweeps up the stellar wind, as expected. Hence, the forward shock is in the stellar atmosphere, fully radiative, and completely unable to accelerate relativistic particles. Only the reverse shock (RS) is suitable for the task. As $r_{\rm{*}} \approx R_{*}$, the magnetic field at the CD is $B_{\rm CD}\approx B_*$. 

The cooling length of the RS is greater than the size of the acceleration region in all cases (see Table \ref{tab: parametros resultados}); this is why the shock is adiabatic and the acceleration efficiency of the process is relatively high: $ \eta_{\rm ac}\sim 10^{-2}$ (see Sect. \ref{sect: acceleration}). The shock velocity is $\approx4.4\times 10^9 \ {\rm cm \ s^{-1}}$ and the temperature of the shocked gas reaches $\approx4.8\times 10^{10} \ {\rm K}$.

We calculate the energy gain and losses of the shock-accelerated  particles following Sect. \ref{sec: radiative processes}. Highly relativistic protons escape from the acceleration region without cooling in all scenarios considered here (with energies up to $E_{\rm p} \approx 1 \ {\rm PeV}$) and are injected into the interstellar medium (ISM). Protons are advected, that is, they are removed from the collision region by the bulk motion of the fluid. They therefore do not interact with ambient material at scales similar to that of the system. Electrons cool mainly through IC and synchrotron mechanisms, and reach a maximum energy of $E_{\rm e}\approx 100 \ {\rm GeV}$. To obtain the electron distribution, we solve the transport equation considering only the dominant IC and synchrotron losses, and a power-law injection function with a spectral index of $2.2$ and an exponential cutoff (see Eq. \ref{eq: injection}).

\subsection{Spectral energy distribution}

\begin{figure*}[h]
\begin{multicols}{2}
    \includegraphics[width=\linewidth]{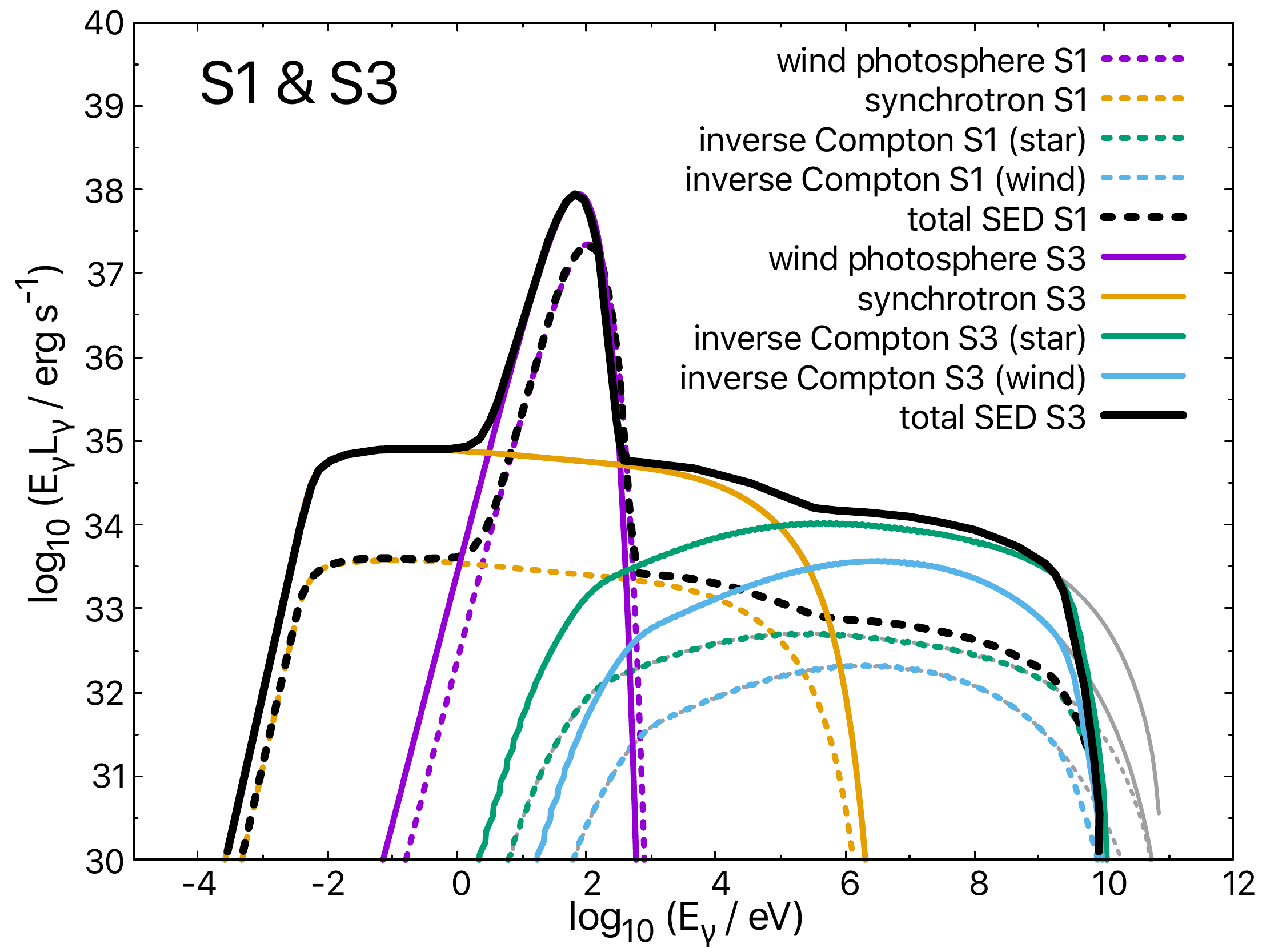}\par 
    \includegraphics[width=\linewidth]{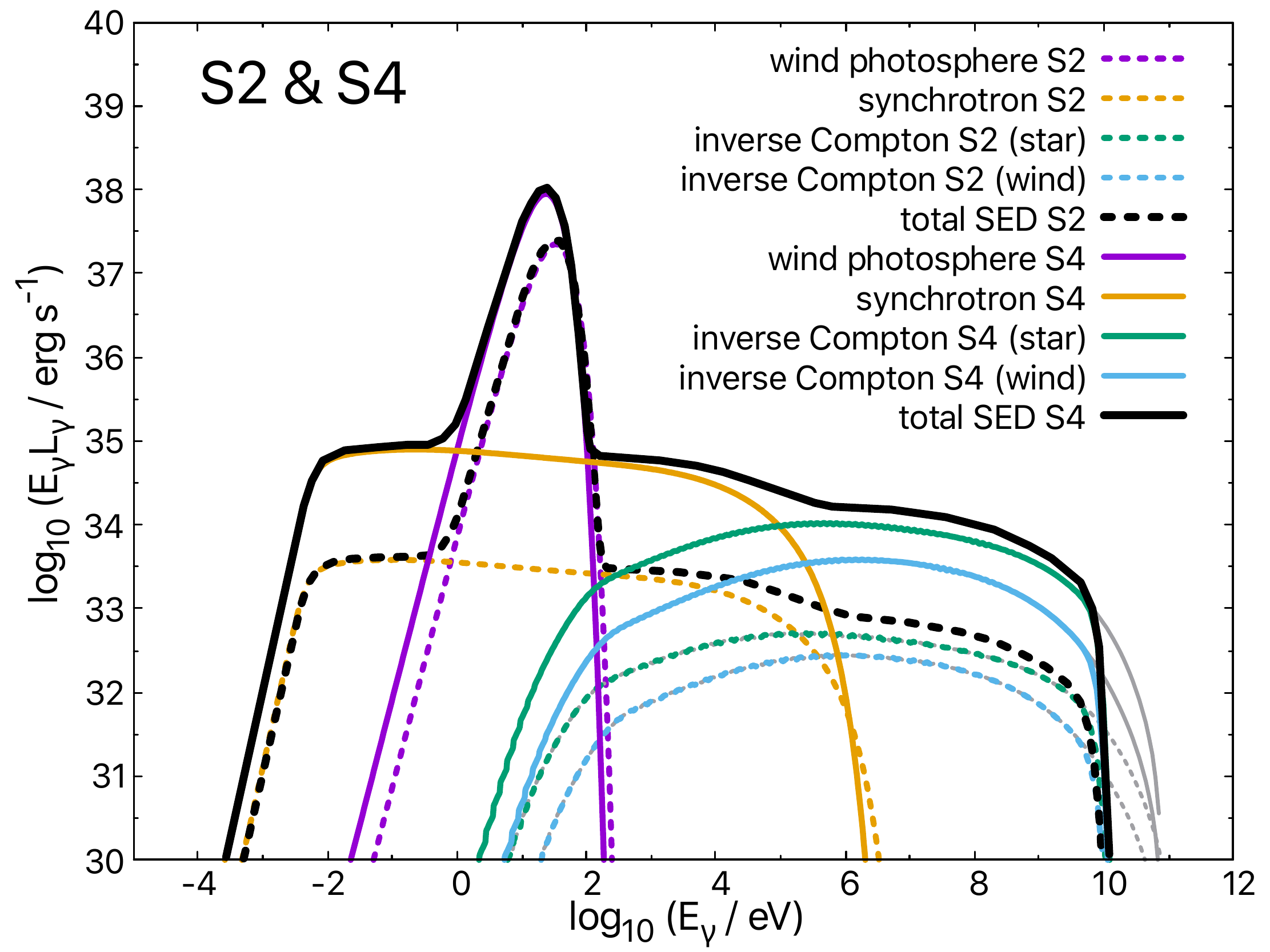}\par 
    \end{multicols}
\caption{Thermal and nonthermal SEDs of the four scenarios considered, S1--S4, in logarithmic scale, where a face-on inclination is assumed. S1 and S3 are shown in the left plot, whereas S2 and S4 are shown in the right plot. Dashed lines correspond to S1 (left) and S2 (right), solid lines correspond to S3 (left) and S4 (right). We plot the nonattenuated inverse Compton contributions  in gray. The emission peak at high energies is $\sim10^{33} \ {\rm erg \ s^{-1}}$ for S1 and S2, and $\sim10^{34} \ {\rm erg \ s^{-1}}$ for S3 and S4. The gamma-ray absorption due to $\gamma\gamma$ annihilation is total for energies > 10 GeV.}
\label{fig: total sed results}
\end{figure*}

 Figure \ref{fig: total sed results} shows the SEDs of the four scenarios. The only thermal component of the spectrum is the photosphere of the optically thick disk-driven wind. The emission peak of the wind for S1 and S2 is $\approx 10^{37} \ {\rm erg \ s^{-1}}$, whereas for S3 and S4 the peak is $\approx 10^{38} \ {\rm erg \ s^{-1}}$. This occurs at energies of $\sim 100$ eV for S1 and S3, and $\sim 30$ eV for S2 and S4. Therefore, if $M_{\rm BH}$ increases, the luminosity is higher and, if the mass-accretion rate increases, the luminosity peak occurs at lower energies.

In the case of the nonthermal spectrum, we calculate the emission due to synchrotron and IC losses. In the latter case, we consider the photon fields of the star and of the wind photosphere as targets. In all cases, the dominant IC contribution is that of the star. The luminosity  in S3 and S4 is an order of magnitude greater than that in S1 and S2. This is because of the modification of the orbital parameters when the BH mass varies: to guarantee the overflow of the Roche lobe, the orbital semi-axis varies with $M_{\rm BH}$, which results in variation in the size of the acceleration region and the photon density at SP, among other parameters. The emission peak at low energies is $\sim 10^{33} \ {\rm erg \ s^{-1}}$ for S1 and S2, and $\sim 10^{35} \ {\rm erg \ s^{-1}}$ for S3 and S4. At high energies, the emission peak is $\sim 10^{32} \ {\rm erg \ s^{-1}}$ (S1 and S2) and $\sim 10^{34} \ {\rm erg \ s^{-1}}$ (S3 and S4). The gamma-ray absorption due to $\gamma\gamma$ annihilation is total for energies > 10 GeV in all scenarios\footnote{We note that, since we assume a nearly face-on inclination of the system, there are no significant variations of the radiative output associated with the orbital phase. If the system were oriented nearly edge-on, the emission would be modulated by the orbital phase due to absorption \citep[for details see][]{2010A&A...518A..12R}.}.

Attenuation due to material between the source and the observer, that is, absorption by external cold gas, is mainly in the optical-to-UV range and at soft X-rays. At radio wavelengths, refractive scintillation on free electrons of the ISM occurs at lower frequencies than predicted here. For high-energy gamma rays, the main absorbers are infrared (IR) fields and the cosmic microwave background (CMB), but their effects are only relevant for cosmological distances.

\section{Application to NGC 4190 ULX 1}\label{sect: ULX1}

Ultraluminous X-ray sources (ULXs) are extragalactic point-like objects where the luminosity in the X-ray band appears to be higher than the Eddington luminosity
\citep{2016AN....337..349B}. ULXs are thought to be X-ray binaries with a stellar-mass compact object accreting at super-Eddington rates, where a beaming effect could be responsible for the luminosity observed in the X-ray band: the radiation emitted from the inner part of the accretion disk is geometrically collimated by the ejected wind, which is optically thick except in a narrow region around the black-hole axis and forms a cone-shaped funnel \citep{2001ApJ...552L.109K,2009MNRAS.393L..41K,2017ARA&A..55..303K,2021AstBu..76....6F}. 

We apply our model to estimate the radiation emitted by the ultraluminous X-ray source NGC 4190 ULX 1 (also known as CXO J121345.2+363754). Although many characteristics of this ULX remain poorly understood, 
several authors have explored the system and have provided constraints on some of its parameters \citep[see e.g.,][]{2005ApJS..157...59L,2013ApJS..206...14G,2017A&A...608A..47K,2018MNRAS.473.5680K,2021MNRAS.504..974G}.

In what follows, we describe the parameterization of the system and its components, and investigate the expected collision of winds. The complete set of parameters used in this section is detailed in Table \ref{tab: parametros generales}.

\subsection{System parameterization}

The source is located in the nearby Galaxy NGC 4190 at a distance of $d\approx3 \ \rm Mpc$ \citep{2013AJ....146...86T}. Observations made in 2010 using the \textit{XMM-Newton} telescope 
reveal a long-term spectral variability in the 0.3--10.0 keV energy range: $L_{\rm X}\sim 3-8\times10^{39}\ \rm erg \ s^{-1}$.  

The angle $i$ between the line of sight and the $z$-axis at which the disk of a ULX is observed determines the components of its spectrum: blackbody disk (BB) or Comptonization. If $i$ is small, the observer is able to look into the funnel and see the innermost part of the disk: the spectrum shows only the BB component, which corresponds to thermal emission of the disk. This type of spectrum is called broadened disk (BD). If $i$ is sufficiently large, another effect is observed: the interaction between photons and wind particles near the disk surface induces a Comptonization that produces a hardening in the spectrum. Most ULXs exhibit a combination of both phenomena in their X-ray spectrum.

\cite{2021MNRAS.504..974G} investigated the spectral properties of NGC 4190 ULX 1 and suggested that the ULX is in a BD state, and that the compact object is a BH with mass $\sim 10-30 M_{\odot}$ accreting at super-Eddington rates. We fit the \textit{XMM-Newton} observations (Epoch 3) with the supercritical advection-dominated disk model detailed in Sect. \ref{subsec: accretion disk}, assuming a mass-accretion rate of $\dot{M}_{\rm input}=10\dot{M}_{\rm Edd}$. We also assume a face-on inclination $i\approx 0^{\circ}$, a BH mass $10M_{\odot}$ and a geometrical beaming factor $b=0.07$.  
This factor is given by,
\begin{equation}
    b=\Omega/4\pi=0.5 (1-\cos{\vartheta}),
\end{equation}
where $\Omega$ is the solid angle of the emission. The angle $\vartheta$ is related to the opening angles of the disk ($\delta$) and its wind (${\theta}$): $\vartheta+\delta+2\theta=90^{\circ}$. 
Both angles, $i$ and $\vartheta$, can change over time, causing the spectral variability of the object \citep{2021AstBu..76....6F}.

On the other hand, \cite{2013ApJS..206...14G} provided constraints on the characteristics of the optical counterpart of the system. They suggested that, if $M_{\rm BH}=10M_{\odot}$, the mass of the star could be $<50 M_{\odot}$ and its radius $<86 R_\odot$. We choose a star of type B2V  for our model in light of one of the fittings these latter authors made from \textit{Hubble Space Telescope} observations. If we apply Eq. \ref{eggleton} and consider the mass ratio $M_*/M_{\rm BH}$, and the stellar radius involved (see Table \ref{tab: parametros generales}), the transfer of mass in the binary system occurs for an orbital semi-axis $a \le 15.2 \, R_{\odot}$, which results in a period $\le 38 \, {\rm h}$.

\begin{table} 
\begin{center}
\caption{Parameters of NGC 4190 ULX 1.}
\label{tab: parametros generales}
\begin{adjustbox}{max width=\columnwidth}
\begin{tabular}{l c c c}
\hline
\hline
\rule{0pt}{2.5ex}Parameter & Symbol & Value & Units  \\
\hline
\rule{0pt}{2.5ex}System \\  
\hline 
\rule{0pt}{2.5ex}Inclination$^{(1)}$ & $i$ & 0 & $^{\circ}$ \\
Orbital semi-axis$^{(2)}$ & $a$ & 15 & $R_{\odot}$ \\
Distance to the source$^{(3)}$ & $d$ & 3 & ${\rm Mpc}$ \\
\hline
\rule{0pt}{2.5ex}Black hole\\  
\hline
\rule{0pt}{2.5ex}Mass$^{(1)}$ & $M_{\rm BH}$    & 10 & $M_{\odot}$ \\
Gravitational radius$^{(2)}$ & $r_{\rm g}$   & $1.48\times 10^6$  & $\rm{cm}$ \\
\hline
\rule{0pt}{2.5ex}Accretion disk\\
\hline
\rule{0pt}{2.5ex}Disk semi opening angle$^{(1)}$ & $\delta$ & 30 & $^{\circ}$ \\
Critical radius$^{(2)}$ & $r_{\rm crit}$ & $3.5\times 10^9$ &  ${\rm cm}$ \\
Eddington accretion rate & $\dot{M}_{\rm Edd}$ & $2.2\times 10^{-7}$ & $M_{\odot} \ \rm{yr}^{-1}$ \\
Mass accretion rate$^{(1)}$ & $\dot{M}_{\rm input}$ & $2.2\times 10^{-6}$ & $M_{\odot} \ \rm{yr}^{-1}$ \\
Mass loss in winds$^{(1)}$ & $\dot{M}_{\rm dw}$ & $1.98\times 10^{-6}$ & $M_{\odot} \ \rm{yr}^{-1}$ \\
Wind velocity$^{(2)}$ & $v_{\rm dw}$ & $4.95\times 10^{9}$ & ${\rm cm \ s^{-1}}$ \\
Wind semi opening angle$^{(2)}$ & $\theta$ & 14.5 & $^{\circ}$ \\
Beaming factor$^{(2)}$ & $b$ & 0.07 & $-$ \\
\hline
\rule{0pt}{2.5ex}B2V Star \\  
\hline
\rule{0pt}{2.5ex}Mass$^{(4)}$ & $M_*$    & 8 & $M_{\odot}$ \\
Radius$^{(4)}$ & $R_*$   & 5.4  & $R_{\odot}$ \\
Temperature$^{(4)}$ & $T_{\rm eff}$    & $20600$ & K  \\
Mass loss in winds$^{(4)}$ & $\dot{M}_*$   & $1.4\times10^{-7}$ & $M_{\odot}\ \rm yr^{-1}$\\
Wind velocity$^{(4)}$ & $v_{\rm *w}$ & $7\times 10^7$ &  $\rm cm \ s^{-1}$ \\
Rotation velocity$^{(1)}$ & $v^{\rm rot}_*$ & $7\times10^6$ &  $\rm cm \ s^{-1}$ \\
Magnetic field$^{(5)}$ & $B_*$ & 200 & G \\
\hline
\rule{0pt}{2.5ex}Colliding winds \\  
\hline 
\rule{0pt}{2.5ex}Kinetic power of disk-driven wind$^{(2)}$ & $L^{\rm dw}_{\rm K}$ & $1.5\times 10^{39}$ & ${\rm erg \ s^{-1}}$ \\
Kinetic power of stellar wind$^{(2)}$ & $L^{*}_{\rm K}$ & $2.17\times10^{34}$ & ${\rm erg \ s^{-1}}$ \\
Distance from BH to SP$^{(2)}$ & $r_{\rm BH}$ & $6.68\times10^{11}$ & ${\rm cm}$ \\
Size of acceleration region$^{(1)}$ & $\Delta x_{\rm ac}$ & $6.68\times10^{10}$ & ${\rm cm}$ \\
Magnetic field at SP$^{(2)}$ & $B_{\rm SP}$ & $200$ & ${\rm G}$ \\
Injection spectral index$^{(1)}$ & $p$ & $2.2$ & $-$ \\
Acceleration efficiency$^{(2)}$ & $\eta_{\rm ac}$ & $10^{-2}$ & $-$ \\
Molecular mean weight$^{(1)}$ & $\mu$ & $0.6$ & $-$ \\
\hline
\rule{0pt}{2.5ex}Reverse shock \\  
\hline
\rule{0pt}{2.5ex}Velocity$^{(2)}$ & $v_{\rm RS}$ & $4.4\times10^9$ & ${\rm cm \ s^{-1}}$ \\ 
Temperature$^{(2)}$ & $T_{\rm RS}$ & $10^{10}$ & ${\rm K}$ 
\\
Cold matter density$^{(2)}$ & $n_{\rm RS}$ & $6.9 \times10^{11}$ & ${\rm cm^{-3}}$ \\
Cooling length$^{(2)}$ & $R_{\rm \Lambda}$ & $2.2\times10^{13}$ & ${\rm cm}$ \\
\hline
\end{tabular}
\end{adjustbox}
\end{center}
\footnotesize{\textbf{Notes.} We indicate the parameters we have assumed  with superscript ${(1)}$  and those we have derived  with ${(2)}$. Parameters with superscripts $(3)$, $(4),$ and $(5)$ were taken from \cite{2013AJ....146...86T}, \cite{2019AJ....158...73K}, and \cite{2015MNRAS.454L...1S}, respectively.}
\end{table}

\subsection{Collision of winds}

The terminal velocity of the disk-driven wind is $v_{\rm dw}=4.95\times10^9 \ \rm{cm \ s^{-1}}$, and therefore $L^{\rm dw}_{\rm K}= 1.5\times10^{39} \ {\rm erg \ s^{-1}}$, while $L^*_{\rm K}= 2.17\times10^{34} \ {\rm erg \ s^{-1}}$. The SP is located near the stellar surface and the wind of the disk completely suppresses the stellar wind. We therefore only take into account the reverse shock (RS). As $r_{\rm{*}} \approx R_{*}$, the magnetic field at the CD is $B_{\rm CD}\approx B_*$. 

\begin{figure}[h] 
        \centering
            \includegraphics[width=\columnwidth]{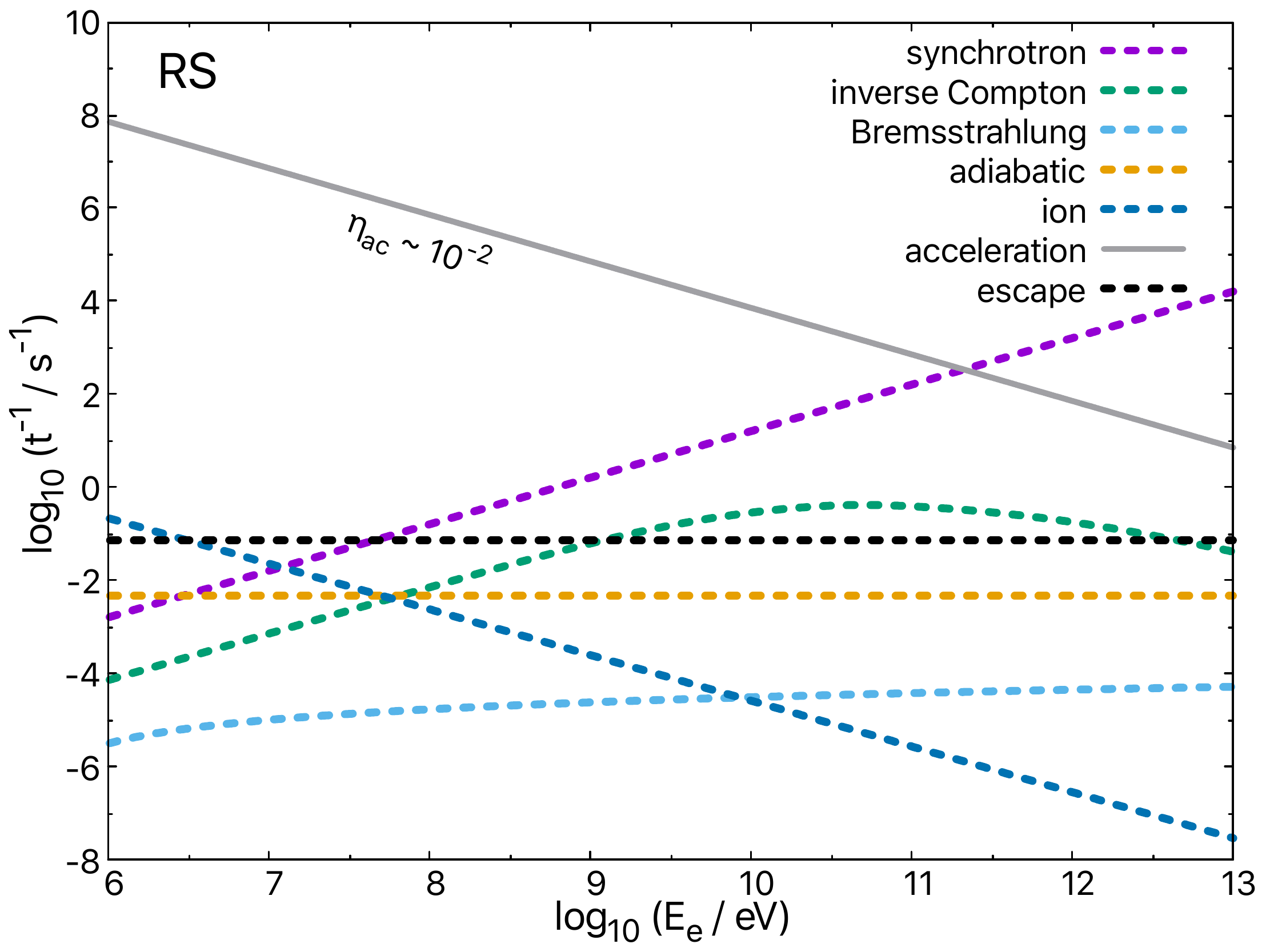}
            \caption{Timescales in logarithmic scale of the electron acceleration, escape, and cooling at the reverse shock in NGC 4190 ULX 1. Electrons reach a maximum energy of $\approx 0.32 \ {\rm TeV}$. The acceleration efficiency is $10^{-2}$.}
\label{fig: times}
\end{figure}

The cooling length of the RS is $R_{\Lambda}=2.2\times10^{13} \ {\rm cm}$ and the size of the acceleration region is $\Delta x_{\rm ac}=6.68\times10^{10} \ {\rm cm}$; therefore, the shock is adiabatic and the acceleration efficiency of the process is $\eta_{\rm ac}=10^{-2}$, as in our general models. We calculate the energy gain and losses of the shock particles following Sect. \ref{sec: radiative processes}. Highly relativistic protons escape from the acceleration region without cooling, as in our previous scenarios (with energies up to $E_{\rm p} \approx 1 \ {\rm PeV}$), and are injected into the ISM.  Electrons cool mainly through IC and synchrotron mechanisms. Figure \ref{fig: times}  shows the timescales of electrons, which reach a maximum energy of $E_{\rm e}\approx 0.32 \ {\rm TeV}$. To obtain the electron distribution, we solve the transport equation taking into account only IC and synchrotron losses, and a power-law injection function with a spectral index of $2.2$ and an exponential cutoff.

\subsection{Total SED}

The SED of the ULX spans a broadband energy range. Figure \ref{fig: total} shows the thermal (wind and accretion disk) and nonthermal (colliding-winds shock) contributions of the system. 
We also show the sensitivity of the instruments ALMA, VLA (sub-mm waves), \textit{Fermi,} and CTA (gamma rays), and observational data from \textit{XMM-Newton}. 

\begin{figure}
    \centering
    \includegraphics[width=\columnwidth]{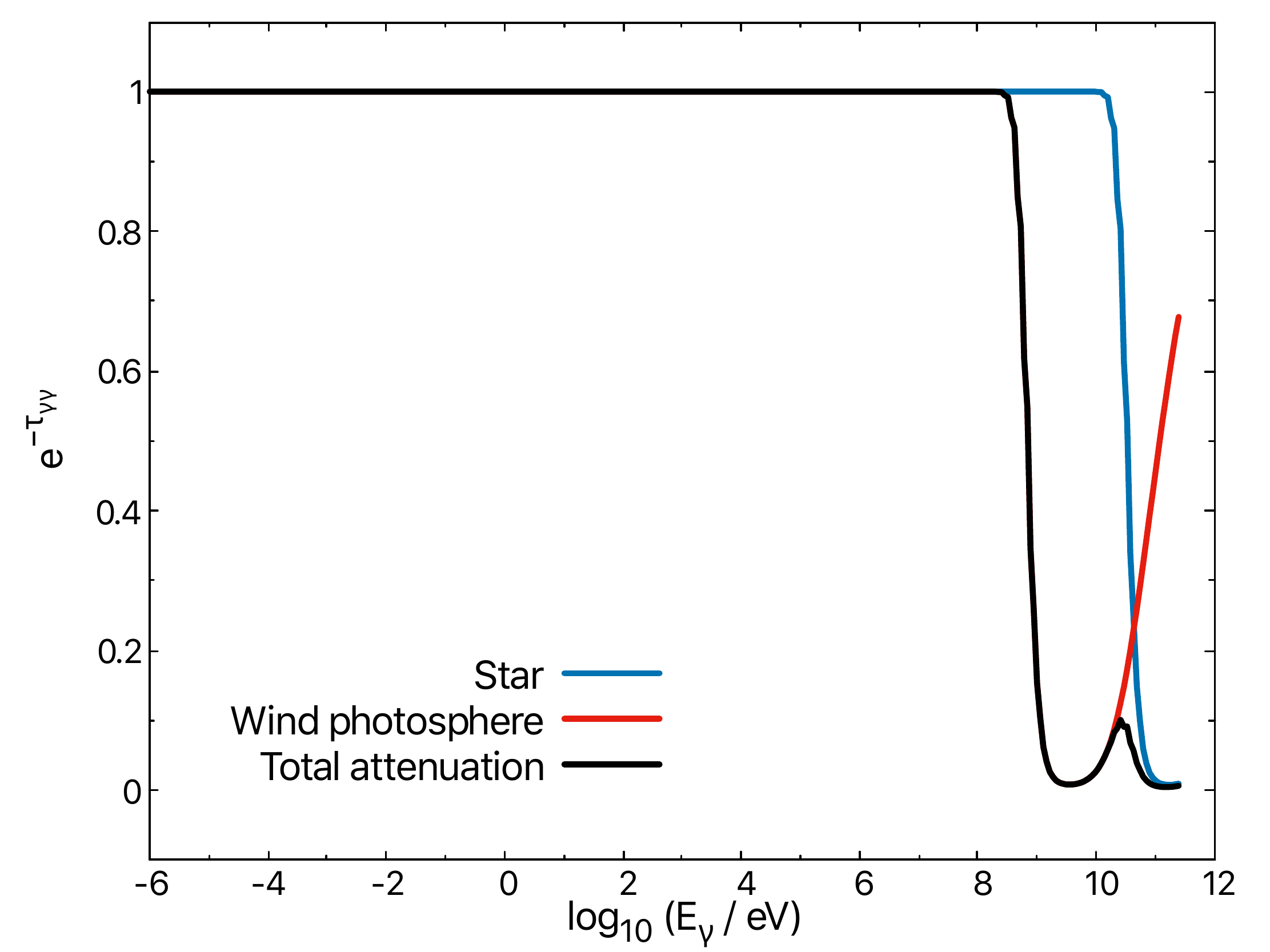}
    \caption{Attenuation factors due to $\gamma \gamma$-annihilation between high-energy nonthermal radiation and photon fields from the star and from the photosphere of the disk-driven wind in NGC 4190 ULX 1. The total attenuation is plotted with a black line.}
    \label{fig: attenuation}
\end{figure}

\begin{figure}[h]
        \centering
            \includegraphics[width=\columnwidth]{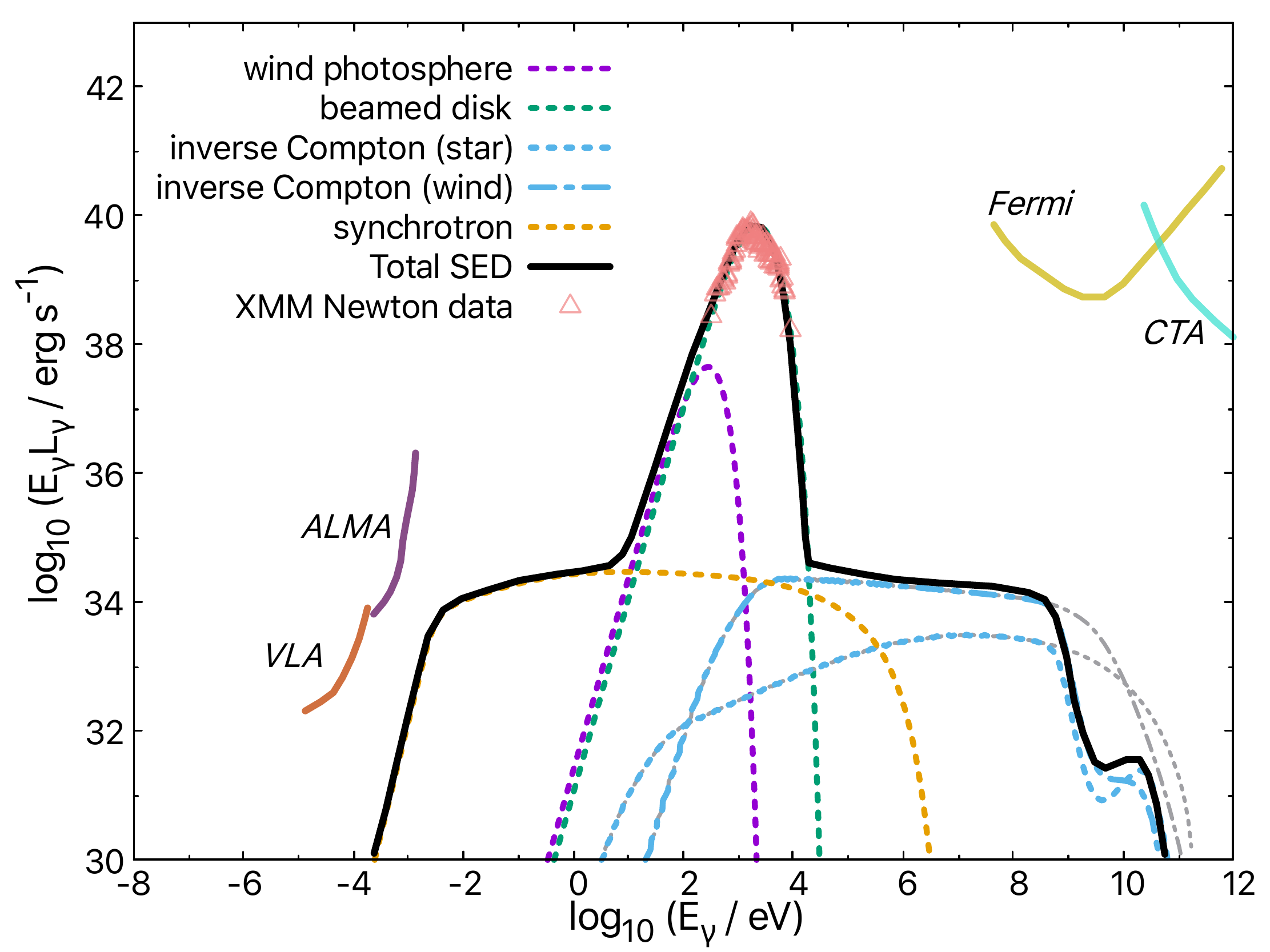}
            \caption{Thermal and nonthermal SEDs of NGC 4190 ULX 1  in logarithmic scale (dashed
lines). The nonthermal SED is partially attenuated for energies > 1 GeV and totally attenuated for energies > 50 GeV due to annihilation of $\gamma$-rays with the photon fields of the star and the photosphere of the disk-driven wind. The gray dashed lines are the nonattenuated IC contributions. The total SED is plotted with a solid black line. Data from \textit{XMM-Newton} (Epoch 3), and the sensitivity of ALMA, \textit{Fermi}, VLA, and CTA are also shown \cite[instrument sensitivities were taken from][]{2022A&A...664A.178S}.}
\label{fig: total}
\end{figure}

The luminosity in the IR  band is $\sim 10^{34} \ {\rm erg \ s^{-1}}$, which is relatively strong, though still undetectable at megaparsec distances. The luminosity in gamma-rays also reaches $\sim 10^{34} \ {\rm erg \ s^{-1}}$. The attenuation factor (Fig. \ref{fig: attenuation}) has an effect on photons with energies $\gtrsim 1 \ {\rm GeV}$. Most of the radiation above 1 GeV and all above 50 GeV is suppressed by the annihilation of the $\gamma$ rays with the photon fields of the disk-driven wind and the star.

\section{Discussion}\label{sect:discussion}

Our analysis of supercritical colliding wind binaries shows that these systems should exhibit broadband emission from radio to gamma rays. In this sense, they are similar to CWBs formed by two hot stars, such as O+WR binaries. However, there are important differences as well. If we compare our models with recent models of O+WR CWBs \citep{2021MNRAS.504.4204P}, we find that (i) in SCWBs, the wind of the disk is far more powerful than the wind of the star. This results in stagnation points that are very close to the surface of the star. Efficient particle acceleration then can only occur in reverse shocks. (ii) We also see that the disk wind advects protons from the acceleration region before they have time to cool. Only electrons can cool locally. The resulting SED is consequently dominated by synchrotron and IC radiation. (iii) As the acceleration region is close to the star, the local magnetic field is relatively strong. Synchrotron emission reaches energies of hundreds of keV. As the medium is far more dense than in stellar CWBs, free-free absorption causes this radiation to turnover below $\sim 24 \, {\rm GHz}$. The total power at millimeter (mm) and submm wavelengths can be between three and five orders of magnitude higher in SCWBs than in stellar CWBs. (iv) IC is the dominant radiation mechanism at high energies. The stronger thermal fields of SCWBs (wind photosphere and star) provide the seed photons, but also impose a high-energy cutoff at $\sim 1$ GeV through $\gamma-\gamma$ attenuation. Instead, stellar CWBs can reach energies close to 1 TeV. (v) The strong magnetic fields in the acceleration region cut electromagnetic cascades in SCWBs. (vi) The SED is always dominated by the X-ray component associated with the disk or its wind in SCWBs. Finally, (vii) stellar CWBs have wider orbits and a variable separation between the components of the system. This produces variability related to the orbital period. On the contrary, the orbits of SCWBs should be mostly circularized. In general, CWBs are weaker than SCWBs, although span a broader energy range.  

An interesting feature of SCWBs is their potential as cosmic ray sources. As mentioned, the strong wind of the disk drags away the relativistic protons before they cool. These protons, with maximum energies of the order of 1 PeV, are then injected into the ISM where they diffuse. Even if a fraction of just $\sim 1$ \% of the wind kinetic power goes to relativistic protons, the cosmic ray output of a SCWB would be in the range $10^{37-39}$ erg s$^{-1}$. These protons might interact with ambient clouds at some distance from the system, producing gamma rays through $pp\rightarrow \pi^0 +pp$ interactions and the subsequent pion decays $\pi^0\rightarrow \gamma\gamma$. The gamma-ray emission from the illuminated clouds can be even stronger than the emission from the binary itself. However, the spectrum should be softer because of propagation effects \citep{1996A&A...309..917A}. Recent modeling by \cite{2021MNRAS.504.4204P} of particle acceleration in colliding wind binaries with wind velocities of a few $10^3$  km $\rm s^{-1}$ and mG magnetic fields in the acceleration region demonstrate that up to $\sim 30$ \% of the wind power can be transferred to nonthermal particles. This means that, in some extreme cases, a SCWB might inject up to $\sim 10^{40}$ erg s$^{-1}$ in cosmic rays.

Another type of CWB is the so-called gamma-ray binary \citep[GRB; e.g., LS 5039, PSR B1259-63, LSI +61$^\circ$ 303, PSR J2032+4127, and others; see, e.g.,][]{2013A&ARv..21...64D,2020mbhe.confE..45C}. These sources are formed by a massive star (usually a Be star with a dense equatorial decretion disk and a fast wind) and a young pulsar in an eccentric orbit. The pulsar ejects a relativistic pair wind. The wind collision produces a broadband spectrum from electrons accelerated at the shock that cool by synchrotron and IC radiation. The two-peak SEDs are similar to those we estimate for SCWBs, but some differences are also clearly seen: (i) GRBs are less energetic because the spin-down luminosity of the pulsar is much smaller than the power of a supercritical wind. (ii) GRBs are highly variable. This variability is modulated with the orbital period. The orbital modulation of the different components of the broadband spectrum is a consequence of the orbital variability of geometrical parameters, such as the geometry of the contact surface of the stellar and pulsar winds. Absorption effects are also strongly variable. (iii) Hadronic interactions are likely when the pulsar crosses the equatorial disk of the star \citep[e.g.,][]{2021ApJ...921L..10B}. (iv) GeV flares have been observed after the periastron passage in sources such as PSR B1259-63 \citep{2011ApJ...736L..11A,2014MNRAS.439..432C}. These flares are attributed to the effects of the unshocked pulsar wind interaction with photons from the stellar disk \citep[e.g.,][]{2012ApJ...752L..17K}.

We finally mention that some black holes accreting at supercritical rates seem to be capable of launching mildly relativistic jets. A remarkable case in our Galaxy is the notorious microquasar SS433 \citep{2004ASPRv..12....1F}. This object resembles a ULX source seen edge on \citep{2006MNRAS.370..399B}. The accretion rate should be extremely high in order to explain the large jet power $L_{\rm K}\sim 10^{40}$ erg s$^{-1}$. \cite{2006MNRAS.370..399B} suggest rates of $\sim 5\times 10^3\; \dot{M}_{\rm Edd}\sim 5\times 10^{-4}\; M_{\odot} \, \mathrm{yr}^{-1}$, which are consistent with estimates of equatorial mass outflows inferred from radio observations \citep{2001ApJ...562L..79B}. These outflows, ejected toward either side of the jets, present a thermal spectrum and might well correspond to the radiation-driven wind of the hypercritical disk. The contamination from the jet base makes it impossible to disentangle contributions from colliding winds from those coming from the jet. However, the equatorial outflow might propagate well beyond the system and reveal itself if it collides with any clouds. The shock generated in the collision would convert the kinetic energy of the plasmoids into internal energy and relativistic particles, which might then cool by $pp$ interactions with the cloud material. Such a scenario might explain the detection of a GeV source by the $Fermi$ satellite on the side of SS433 \citep{2020NatAs...4.1132B,2020NatAs...4.1177L}. We will explore the details of this hypothesis elsewhere.

\section{Summary and conclusions}

We explored the consequences of supercritical accretion in binary systems consisting of a hot star and a black hole.
We find that a fraction of the kinetic power of the radiation-driven wind released by the accretion disk is transformed into relativistic particles in the region of the wind that collides with the star. Electrons are cooled locally, mainly through synchrotron and inverse Compton radiation. The radiation fields of the star and wind photosphere provide abundant thermal photons for the latter process; they also absorb high-energy radiation above a few GeV. Free-free absorption imposes a high-frequency turnover in the radio regime, suppressing centimeter radio waves, unlike the case of colliding wind binaries. The relativistic protons are blown away by the wind before they can cool down significantly. Once trapped by the outflow, these protons are transported to outer regions where they can interact with ambient gas away from the binary system, producing hadronic gamma-rays. Our most important finding is that,  in addition to being strong
thermal UV and X-ray sources, supercritical colliding wind binaries can be significant nonthermal sources at mm wavelengths and GeV energies.

\begin{acknowledgements}
The authors thank the anonymous referee for a careful and constructive review, and for his/her comments that improved this work. We thank also Daniela Pérez and Ji\v{r}i Hor\'ak for fruitful discussions. This work was supported by grant PIP 0554 (CONICET). LA ackowledges the Universidad Nacional de La Plata for the education received. GER acknowledges the support from the Spanish Ministerio de Ciencia e Innovaci\'on (MICINN) under grant PID2019-105510GBC31 and through the Center of Excellence Mara de Maeztu 2020-2023 award to the ICCUB (CEX2019-000918-M).
\end{acknowledgements}

\bibliographystyle{aa} 
\bibliography{biblio} 

\end{document}